\documentclass[twocolumn,showpacs,preprintnumbers,amsmath,showkeys, amssymb]{revtex4}
\usepackage[unicode]{hyperref}
\usepackage[caption=false]{subfig}
\usepackage{graphics}
\usepackage{comment}
\usepackage{epstopdf}
\usepackage{epsfig}
\usepackage{graphicx}
\graphicspath{ {./images/} }
\usepackage{dcolumn}
\usepackage{bm}
\usepackage[caption=false]{subfig}
\usepackage[utf8]{inputenc}
\usepackage{dirtytalk}
\usepackage[autostyle]{csquotes}
\usepackage{xcolor}
\usepackage{scalerel}

\begin{document}

\title{
Percolation on multifractal, scale-free weighted planar stochastic porous lattice
}

\author{Proshanto Kumar and Md. Kamrul Hassan
}
\date{\today}

\affiliation{
Theoretical Physics Group, Department of Physics, University of Dhaka, Dhaka 1000, Bangladesh
}

\begin{abstract}%

We introduce the Weighted Planar Stochastic Porous Lattice (WPSPL), a geometrically disordered substrate generated by iteratively subdividing a unit square. At each step a block is selected with probability proportional to its area, divided into four parts, and one sub-block is retained (removed) with probability $q$ ($1-q$). We show analytically that the WPSPL exhibits multifractality for each of its infinitely many nontrivial conserved quantities and demonstrate numerically that its snapshots at different times are statistically self-similar. The dual of the lattice forms a complex network with a power-law degree distribution. Motivated by these properties of this porous lattice, we study bond percolation on the WPSPL, determine the percolation threshold, and estimate the critical exponents $\alpha$, $\beta$, and $\gamma$ associated with the specific heat, order parameter, and susceptibility, respectively. The exponents vary continuously with $q$, reflecting a family of distinct universality classes as the global dimension of the lattice depends on $q$. Remarkably, the Rushbrooke inequality, $\alpha + 2\beta + \gamma \ge 2$, is satisfied in near equality. Notably, the nonporous case ($q=1$) has a global dimension $2$ but lies outside the universality class of conventional two-dimensional lattices. Our results highlight how geometric disorder, multifractality, scale-free coordination number disorder, and porosity produce unconventional critical behavior.

\end{abstract}



\maketitle

\section{Introduction}

Percolation theory provides a minimal yet powerful framework for understanding the emergence of large-scale connectivity in disordered systems. Its origins date to the early 1940s, when Flory introduced percolation concepts in the context of polymer gelation \cite{ref.flory}. A rigorous mathematical formulation was established by Broadbent and Hammersley in 1957 to describe fluid flow through porous media \cite{ref.broadbent}, thereby formalizing the role of randomness and geometry in connectivity transitions. During the 1960s and 1970s, percolation theory emerged as a paradigmatic model of continuous phase transitions through the seminal contributions of Domb, Fisher, Essam, and Sykes \cite{DombSykes1957,Fisher1967, SykesEssam1964, ref.essam1980}. This period saw the development of key concepts such as finite-size scaling, renormalization group ideas, and universality, firmly embedding percolation within the modern theory of phase transition and critical phenomena \cite{ref.KADANOFF,ref.Wilson_1,ref.Wilson_2,ref.Wilson_3,ref.barbar,ref.Shang}. Subsequent advances in the 1980s and 1990s by Aharony, Stauffer, Ziff, and others further broadened its scope and applications, establishing percolation as a unifying framework for disordered media, transport processes, and complex networks \cite{ref.Stauffer,ref.Sahimi,ref.Ziff,BundeHavlin1996,Stanley1999}.

Apart from providing a theoretical framework for understanding phase transitions and critical phenomena, percolation theory has found applications in a wide range of fields, including epidemic spreading, wildfire dynamics, network robustness, and the propagation of biological, computer, and information contagions in social and technological networks \cite{ref.Newman_virus,ref.Cohen_virus,ref.Moore_virus,ref.opinion_1,ref.boccaletti_opinion}. Recent studies further demonstrate that percolation transitions govern macroscopic behavior in diverse physical systems, such as transport near the percolation threshold in porous media and mechanically induced connectivity in granular matter \cite{Residori2025,KimWuHan2025}. The COVID-19 pandemic has underscored the relevance of percolation-based models for capturing the complexity of real-world spreading processes \cite{ref.epidemic_vespignani,ref.infection,ref.sir,ref.barabasi_decade,ref.Cattuto}. Beyond these contexts, percolation concepts have also found applications in high-energy physics and cosmology, where they help elucidate deconfinement transitions through the color string percolation model and the emergence of large-scale cosmic connectivity \cite{ref.CSPM,ref.cosmicweb}.
   
Any percolation model is defined by two essential ingredients: the percolation rule and the underlying substrate \cite{ref.broadbent,ref.flory}. Until recently, most studies focused on regular spatial lattices. Such lattices provide appropriate models for crystalline solids. However, they fail to capture many real spreading processes. Examples include epidemics, information flow, and transport in porous media. These processes typically occur on substrates that are disordered, heterogeneous, and dynamically evolving \cite{ref.Newman_virus,ref.Cohen_virus}. This mismatch has motivated the development of spatially embedded models with scale-free, self-similar, and complex geometric structures \cite{ref.boccaletti_opinion}. In this context, Hassan \textit{et al.} introduced the \emph{Weighted Planar Stochastic Lattice} (WPSL) in 2010 and showed that, despite its intrinsic geometric disorder, it exhibits remarkably robust statistical regularities \cite{ref.hassan_njp,ref.hassan_jpc}. Most notably, its configurations are statistically self-similar across growth stages, the coordination-number distribution follows a power law, and the growth dynamics are governed by infinitely many conservation laws, giving rise to multifractal behavior.

In 2015, we studied percolation on the Weighted Planar Stochastic Lattice (WPSL) and showed that, although it is two-dimensional, it does not belong to the universality class of regular two-dimensional lattices \cite{ref.hassan_njp,ref.hassan_jpc}. Consistent with general percolation theory, however, site and bond percolation on the WPSL share the same universality class, and the associated critical exponents were later shown to satisfy the Rushbrooke inequality \cite{ref.hassan_njp,hassan2017entropy}. These results established the WPSL as a physically meaningful substrate for studying critical phenomena on disordered, scale-free spatial lattices. More recently, we introduced a stochastic porous variant of the WPSL in which each selected block is subdivided into two rectangles, either horizontally or vertically \cite{mitra2021multi}. One of the two newly generated blocks is retained (removed) with probability $q$ ($1-q$). Analytical results reveal an infinite set of conserved quantities, each defining a multifractal measure supported on a porous fractal substrate of dimension $2q$ for $0<q<1$. When blocks are characterized by their areas, the resulting block-size distribution is found to obey dynamic scaling \cite{mitra2021multi}.

Apart from demonstrating that the WPSPL is a multifractal, scale-free, and self-similar lattice, the main goal of this article is to study percolation on the weighted planar stochastic porous lattice (WPSPL), a model that captures the intrinsic disorder and stochastic geometry of porous media. The porosity, controlled by $q$, strongly influences global connectivity. In particular, the percolation threshold $p_c$ increases with $q$, indicating reduced connectivity at higher porosity. Employing a thermodynamic analogy, we interpret the occupation probability $p$ as an external ordering field and $1-p$ as an effective temperature. This mapping enables us to define percolation analogs of susceptibility and specific heat. Using finite-size scaling, we then estimate the critical exponents $\alpha$, $\beta$, and $\gamma$. Although these exponents vary continuously with $q$, they satisfy the Rushbrooke inequality $\alpha + 2\beta + \gamma \ge 2$ with near equality for all $q$, consistent with the static scaling hypothesis \cite{stanleyBook}. Consequently, for each value of $q$, percolation on the WPSPL defines a distinct universality class, different from that of regular planar lattices.


The paper is organized as follows. In Sec.~II, we introduce the weighted planar stochastic porous lattice (WPSPL) by incorporating probabilistic mass loss into the WPSL and describe the construction procedure and the underlying algorithm. In Sec.~III, we present an analytical treatment of the model and perform a multifractal analysis, demonstrating that the lattice is governed by infinitely many multifractal measures. In Sec.~IV, we establish the statistical self-similarity of the lattice across different system sizes and growth stages. Section~V contains the main results of this work, where we study bond percolation on the WPSPL, determine the percolation threshold, extract the associated critical exponents, and discuss the resulting universality classes and their dependence on the porosity parameter $q$. Finally, a summary and discussion are provided in Sec.~VI.

\section{Construction of WPSPL}

In 2010, Hassan \textit{et al.} introduced the \emph{weighted planar stochastic lattice} (WPSL), a space-filling structure generated by random sequential partitioning of the plane into contiguous, non-overlapping blocks \cite{ref.hassan_njp}. Starting from a unit-area square, the initiator is first divided into four blocks. Subsequently, at each step, a block is selected with probability proportional to its area and subdivided by two mutually perpendicular cuts into four smaller blocks. This construction, known as WPSL2, is intrinsically disordered yet exhibits robust statistical regularities. Notably, its coordination-number distribution follows an inverse power law, implying that the dual network is scale-free with degree distribution $P(k)\sim k^{-\gamma}$ and $\gamma=5.66$ \cite{ref.hassan_njp}. Moreover, the dynamics are governed by infinitely many conservation laws, each generating a multifractal measure, establishing WPSL2 as a multi-multifractal system \cite{ref.hassan2011, ref.dayeen2016}.

\begin{figure}
\centering

\includegraphics[width=8.5cm,height=8.5cm,clip=true]
{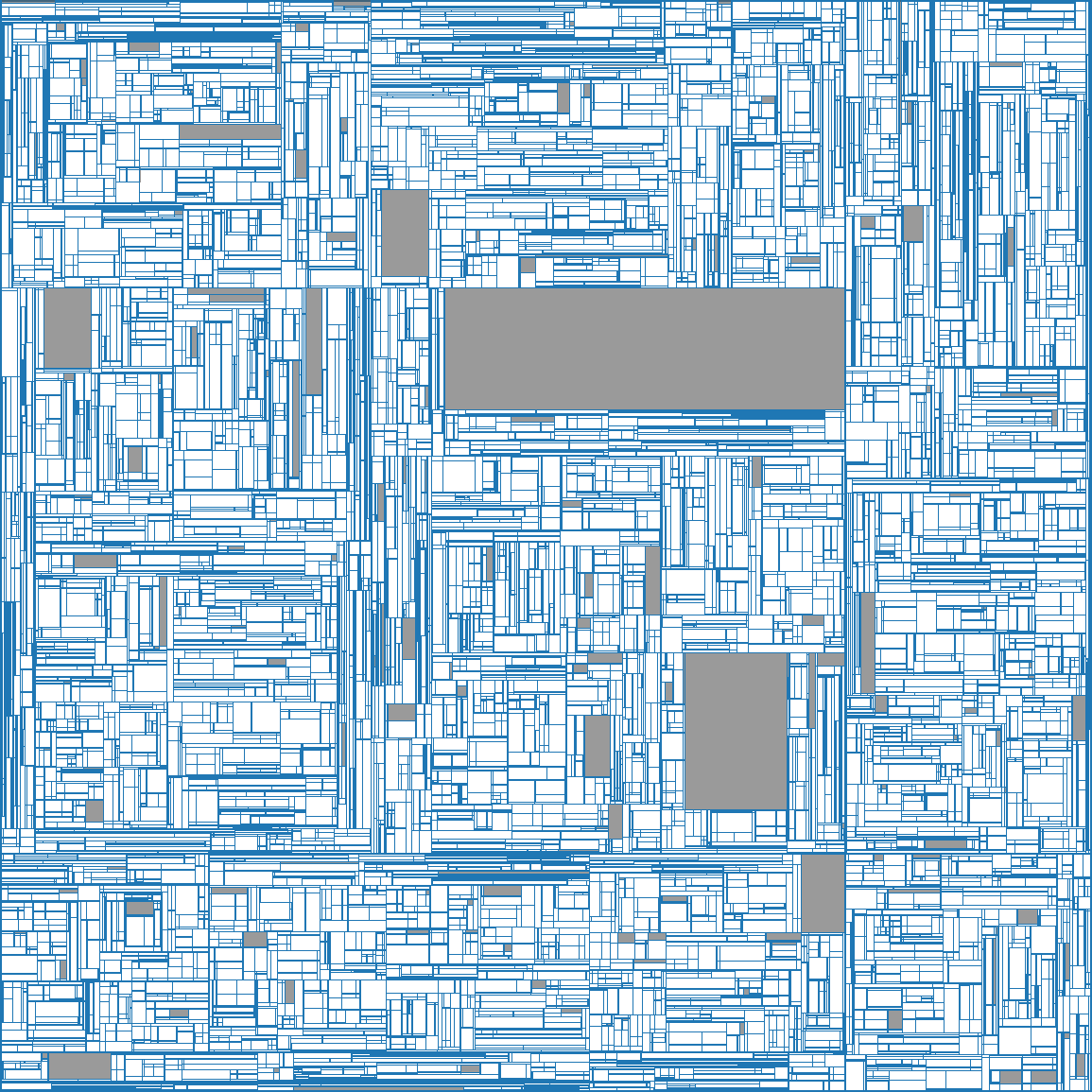}
\label{fig:1}

\caption{A snapshot of the stochastic for p=0.95 and t=5000. The shaded cells indicate that the cells were deleted.
}

\label{fig:1}
\end{figure}

We briefly describe the construction of the \emph{weighted planar stochastic porous lattice} (WPSPL). The WPSPL is generated following the same stochastic subdivision protocol as the WPSL, with the additional rule that porosity is introduced via probabilistic block removal. Starting from a unit-area square, each selected block is subdivided by two mutually perpendicular cuts into four smaller blocks. Of these, one designated block (the choice is conventional and does not affect observables) is retained with probability $q$ or removed with probability $1-q$, creating a void. Block areas $a_i$ serve as selection weights and satisfy the normalization $\sum_{i=1}^{4} a_i = 1$. At each subsequent step, a block (or void) is chosen with probability proportional to its area. If a void is selected, no subdivision occurs and time is incremented by one; otherwise, the subdivision and retention procedure is repeated. This process generates a stochastic porous lattice with tunable porosity controlled by $q$.

We describe the generalized $j$th step in the construction of the weighted planar stochastic porous lattice (WPSPL). At step $j-1$, the system consists of a collection of planar blocks, each characterized by its area and state (occupied or void).

\begin{itemize}

\item[(i)] \textbf{Block Selection.}
\begin{itemize}
    \item[(a)] Consider the set of areas of all surviving (non-void) blocks $\{a_1,a_2,\ldots,a_N\}$,
    where $N$ is the number of surviving blocks. Let $M=\sum_{k=1}^{N} a_k \le 1$
    denote the total occupied area.

    \item[(b)] Generate a random number $R_1$ uniformly distributed in the interval $[0,1]$.

    \item[(c)] If $R_1 \le M$, select a block according to its area by finding the smallest integer $n$ such that
    $\sum_{k=1}^{n-1} a_k < R_1 \le \sum_{k=1}^{n} a_k$.    The $n$th block, of area $a_n$, is then selected for subdivision and go to step (ii).

    \item[(d)] If $R_1 > M$, the selected entity is a void. In this case, no subdivision is performed, the configuration remains unchanged, and the system time is increased by one unit and go to step (i)
\end{itemize}

\item[(ii)] The above procedure ensures that an occupied block $i$ of area $a_i$ is selected with probability
\[
\Pi_i=\frac{a_i}{\sum_{k} a_k},
\]
i.e., block selection is preferential with respect to area.

\item[(iii)] \textbf{Subdivision of an Occupied Block.}
If a surviving block is selected, apply the generator to subdivide it into four smaller blocks. This is achieved by choosing a point uniformly at random inside the block and drawing two mutually perpendicular lines through this point, parallel to the sides of the block. The subdivision produces four contiguous sub-blocks of arbitrary areas, which are labeled according to a fixed but arbitrary convention (e.g., clockwise starting from the top-left corner).

\item[(iv)] \textbf{Retention Test (Porosity Rule).}
One of the four newly created sub-blocks (chosen by convention) is subjected to a retention test. Generate a random number $R_2$ uniformly distributed in $[0,1]$ and compare it with the retention probability $q$. If $R_2 \le q$, the block is retained; otherwise, it is discarded (converted into a void) with probability $1-q$. The remaining three sub-blocks are always retained.

\item[(v)] \textbf{Configuration Update.}
The parent block is removed from the configuration and replaced by the four daughter blocks, each assigned its corresponding area and state (occupied or void). The labels of the list of surviving blocks and their areas is updated accordingly.

\item[(vi)] \textbf{Iteration.}
Increment the step counter $j \rightarrow j+1$ and repeat the procedure.

\end{itemize}

The labeling of sub-blocks is purely conventional and has no physical significance, as it does not affect any observable. A representative snapshot of the lattice Fig.~(\ref{fig:1}) illustrates the irregular coexistence of retained blocks and voids that emerges during growth. As the lattice evolves in time $t$, the number of blocks $N(t)$ increases, while the total mass (area) decreases for $0<q<1$ due to probabilistic block removal; only for $q=1$ is total area conserved. Despite this strong spatial and temporal disorder, the lattice exhibits statistical self-similarity, manifested through multifractality and dynamic scaling of the block-area distribution. These properties ensure self-similarity in both space and time, which is essential for studying percolation at different lattice sizes and growth stages.

\section{Multifractal properties of stochastic porous lattice}

In this section, we analytically address aspects of the model by employing the kinetics of planar fragmentation to understand the influence of block sizes when size is treated as a dynamical variable \cite{ref.rodgers1994, ref.krapivsky2010, ref.hassan1996}. We characterize the remaining blocks by their length $x$ and width $y$, and define the block-size distribution $f(x,y,t)$ such that $f(x,y,t)dxdy$ gives the number of blocks with lengths and widths in $[x,x+dx]$ and $[y,y+dy]$. Its evolution obeys the master equation
\begin{eqnarray} 
\label{eq:WPSL} 
{{\partial f(x,y,t)}\over{\partial t}} & =& -xy f(x,y,t)+(3+q) \\ \nonumber &\times & \int_x^\infty\int_y^\infty f(x_1,y_1,t)dx_1dy_1. 
\end{eqnarray}
The first term represents the loss of blocks of size $(x,y)$ due to breakup, while the second term accounts for the gain from blocks with $(x_1>x, y_1>y)$. The factor $(3+q)$ reflects that, out of four newly created rectangles, three are always retained and the fourth is retained with a probability of $q$.

Exact solutions for $f(x,y,t)$ are generally intractable due to the stochastic and geometric complexity of the process. Following Krapivsky and Ben-Naim, we instead consider the two-dimensional Mellin transform
\begin{equation}
M(m,n;t) = \int_0^\infty \int_0^\infty x^{m-1} y^{n-1} f(x,y;t)\,dx\,dy,
\end{equation}
whose discrete counterpart is $\sum_i x_i^{m-1} y_i^{n-1}$ \cite{ref.krapivsky1994}. This transform encodes the global geometric information of the lattice and allows for the extraction of scaling properties without explicit knowledge of $f(x,y,t)$. Substituting this definition into the rate equation governing the evolution of
$f(x,y;t)$, we obtain the following hierarchy of moment equations:
\begin{equation}
\frac{dM(m,n;t)}{dt}
=\left(\frac{3+q}{mn}-1\right)M(m+1,n+1;t).
\end{equation}
A remarkable consequence of this equation is the existence of infinitely many
nontrivial conservation laws: all moments of the form $M(m,(3+q)/m;t)$ remain
independent of time for arbitrary $m$. These conserved quantities immediately
suggest a deep underlying multifractal structure.

To solve the moment hierarchy, we iteratively generate all time derivatives of
$M(m,n;t)$ and apply Charlesby’s method by inserting them into the Taylor series
expansion of $M(m,n;t)$ about $t=0$ \cite{ref.charlesby1954}. This procedure yields an exact solution in
terms of generalized hypergeometric functions,
\begin{equation}
M(m,n;t)= {}_2F_2\!\left(a_+,a_-;m,n;-t\right),
\end{equation}
which, in the long-time limit, simplifies to the asymptotic form
\begin{equation}
\label{eq:mnt}
M(m,n;t)\sim t^{-a_-},
\end{equation}
where
\begin{equation}
a_-=\frac{m+n}{2}-\sqrt{\left(\frac{m-n}{2}\right)^2+(3+q)}.
\end{equation}
This asymptotic behavior encapsulates the full scaling content of the WPSL.

The conserved measure associated with the invariant moment $M(m,(3+q)/m;t)$ assigns to
the $i$-th block a weight
\begin{equation}
p_i=x_i^{[(3+q)/m]-1}y_i^{m-1} \quad \text{or} \quad x_i^{m-1}y_i^{[(3+q)/m]-1},
\end{equation}
which can be naturally interpreted as a probability. The corresponding partition
function is
\begin{equation}
Z_k=\sum_i p_i^q.
\end{equation}
Expressing this quantity in terms of the Mellin moments, we obtain
\begin{equation}
Z_k= M\!\big([(3+q)/m-1]k+1,(m-1)k+1;t\big).
\end{equation}
Using the asymptotic form of $M(m,n;t)$, the long-time behavior of $Z_q$ follows as
\begin{equation}
Z_k(t)\sim
t^{\frac{1}{2}\left[\sqrt{((3+q)/m-m)^2k^2+16}-((3+q)/m+m-2)k+2\right]}.
\end{equation}
To express this scaling in geometric terms, following the procedure of multifractal analysis by Feder we introduce a natural yardstick
\begin{equation}
\delta(t)=\sqrt{\frac{M(2,2;t)}{M(1,1;t)}}\sim t^{-1/2},
\end{equation}
which represents the square root of the mean block area at time $t$ \cite{feder1988fractals}.

\begin{figure}
\centering
\subfloat[]
{
\includegraphics[height=4.2 cm, width=4.2 cm, clip=true]
{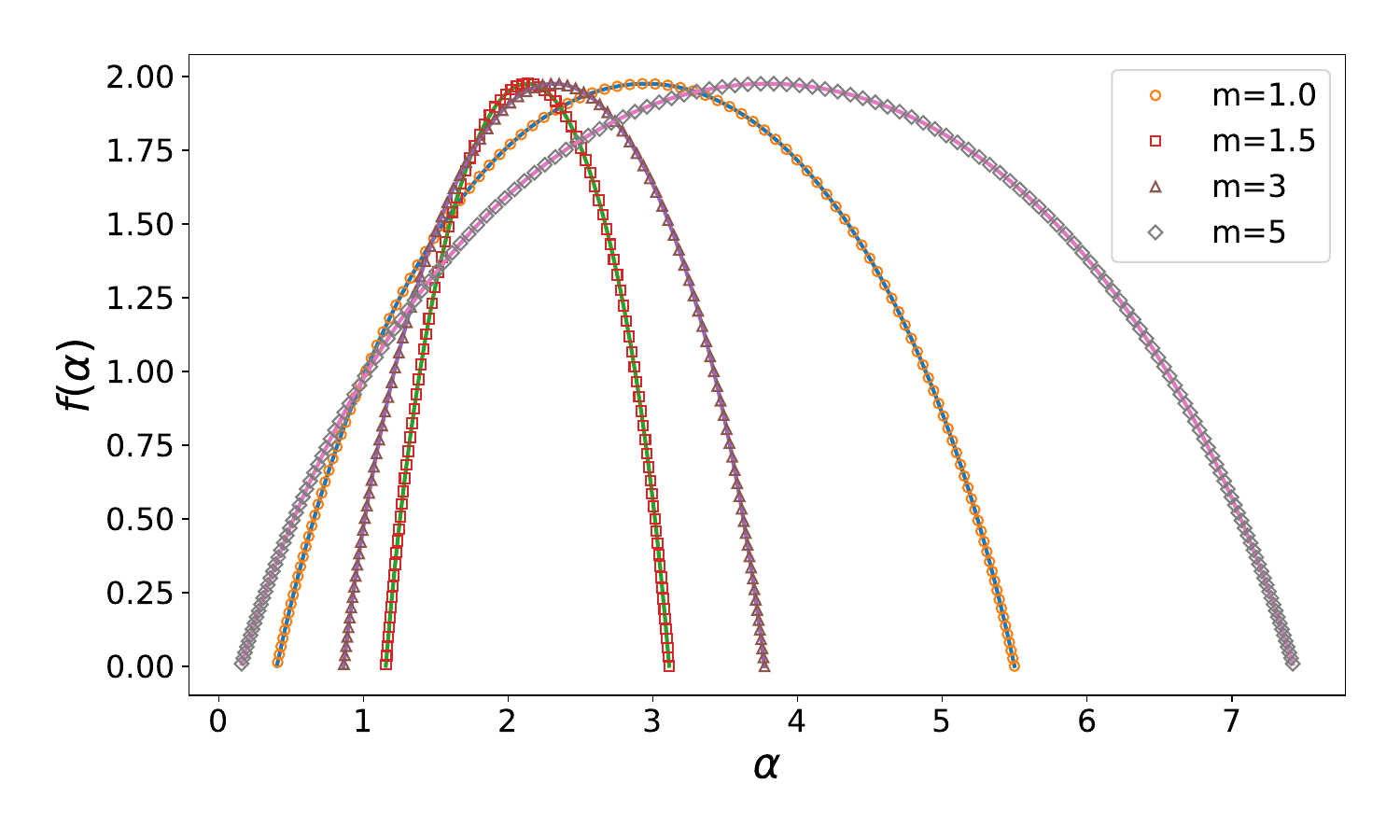}
\label{fig:2a}
}
\subfloat[]
{
\includegraphics[height=4.2 cm, width=4.2 cm, clip=true]
{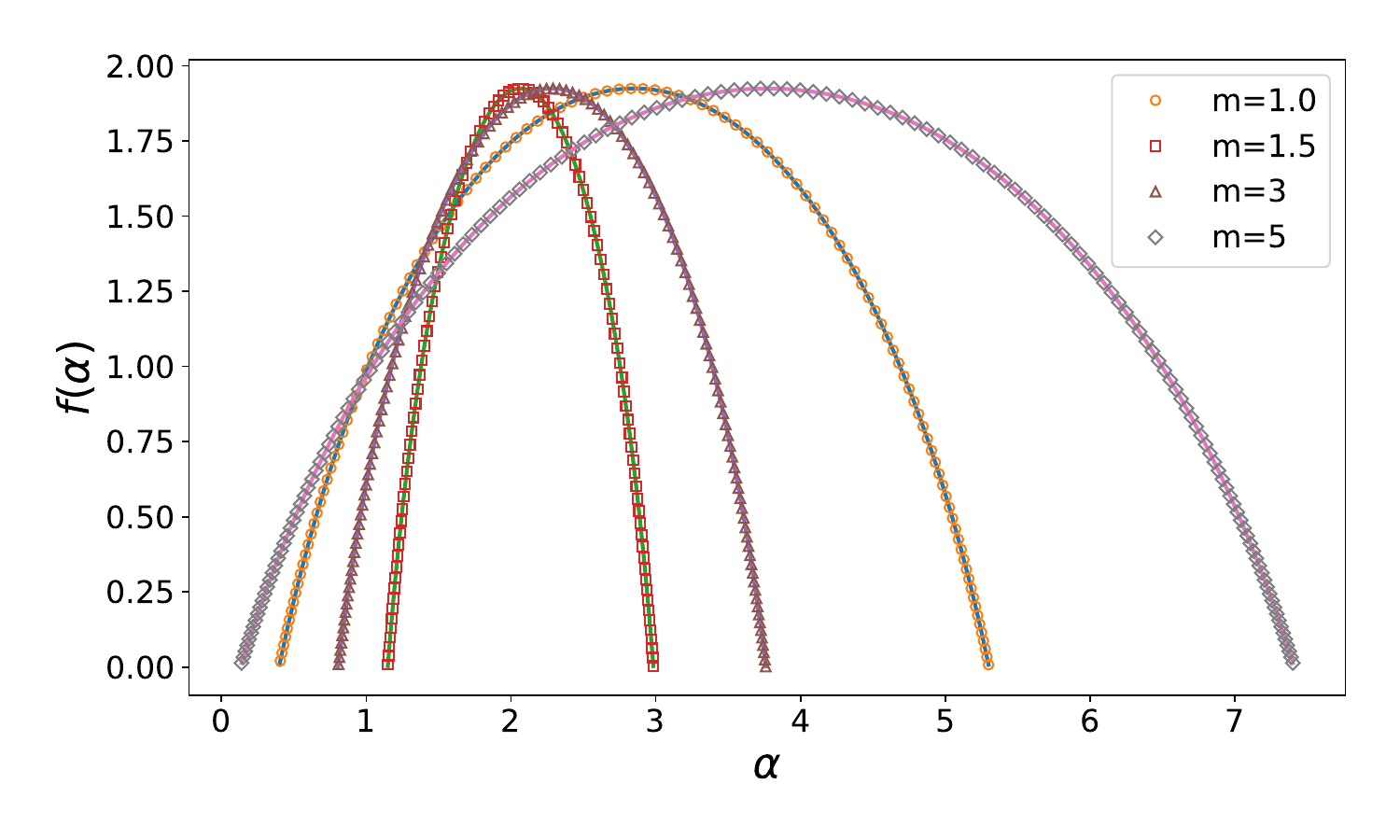}
\label{fig:2b}
}

\caption{ The $f(\alpha(k))$ spectra for (a) $q = 0.95$ and (b) $q = 0.85$, shown for $m = 1,\ 1.5,\ 3,$ and $5$ in each case. In all plots, the maximum occurs at $k = 0$, which corresponds to the fractal dimension of the support, in agreement with the theoretical prediction.} 
\label{fig:2}
\end{figure}

Eliminating time in favor of $\delta$, the partition function assumes the standard multifractal
form
\begin{equation}
Z_k(\delta)\sim \delta^{-\tau(k,m)},
\end{equation}
with the mass exponent
\begin{eqnarray}
\tau(k,m)
&=& \sqrt{((3+q)/m-m)^2k^2+4(3+q)} \\ \nonumber &-& \big(((3+q)/m+m-2)k+2\big).
\end{eqnarray}
The mass exponent satisfies the fundamental multifractal constraints
$\tau(0,m)=2(\sqrt{3+q}-1)$, which is always less than the dimension of the embedding space $2$ for $0<q<1$ and hence the skeleton is a fractal. On the other hand, we see
$\tau(1,m)=0$, as required by probability normalization. Its Legendre transform,
\begin{equation}
\tau(k,m)=-\alpha k+f(\alpha),
\quad \text{where} \quad
\alpha=-\frac{d\tau(k,m)}{dk},
\end{equation}
yields the multifractal spectrum
\begin{equation}
f(\alpha(k),m))
=\frac{16}{\sqrt{((3+q)/m-m)^2k^2+4(3+q)}}-2.
\end{equation}
The spectrum $f(\alpha(k),m)$ is shown in Fig.~\ref{fig:2} for fixed $q$ and several values of $m$. In all cases, the spectra are strictly concave and attain a maximum value $2(\sqrt{3+q}-1)$ at $k=0$, corresponding to the fractal dimension of the WPSPL. Notably, the functional form of $f(\alpha,m)$ depends explicitly on $m$, demonstrating the presence of a family of intertwined multifractal spectra rather than a single universal one.

Multifractality provides a natural framework for understanding scaling and self-similarity in the weighted planar stochastic porous lattice (WPSPL), where disorder is intrinsic and dynamically generated. Unlike regular lattices or simple fractals characterized by a single scaling exponent, the WPSPL evolves through stochastic fragmentation and removal, producing strong spatial heterogeneity in block sizes, masses, and connectivity. Consequently, different regions scale differently under coarse-graining, rendering a single fractal dimension inadequate. In the WPSPL, multifractality describes how local measures—such as block mass, area, or weight—are distributed across multiple length scales. 
Crucially, multifractality explains the statistical self-similarity of the WPSPL. Although individual snapshots appear highly irregular, their statistical properties remain invariant under appropriate rescaling within a single configuration. 

\section{Self-similar properties}

To employ the WPSPL framework for percolation studies, it is first necessary to establish that lattice snapshots of different sizes are self-similar, much like geometrically similar triangles. Figure~\ref{fig:1} illustrates the complex structure that emerges in the long-time limit, where surviving blocks form a heterogeneous pattern interspersed with voids. This naturally raises the question of whether configurations generated at different times are statistically equivalent under appropriate rescaling. In physics, self-similarity can be spatial or temporal. Spatial self-similarity has already been established by demonstrating that the WPSPL is not merely multifractal but exhibits an infinite hierarchy of multifractality. To establish temporal self-similarity, we now show that the block area–size distribution obeys dynamic scaling. Dynamic scaling provides a stringent test of temporal self-similarity and offers a robust framework for identifying scale-invariant behavior in evolving systems \cite{ref.banerjee2019, ref.hassan2011, ref.mitra2021}.

\begin{figure}
\centering
\subfloat[]
{
\includegraphics[height=4.0 cm, width=4.0 cm, clip=true]
{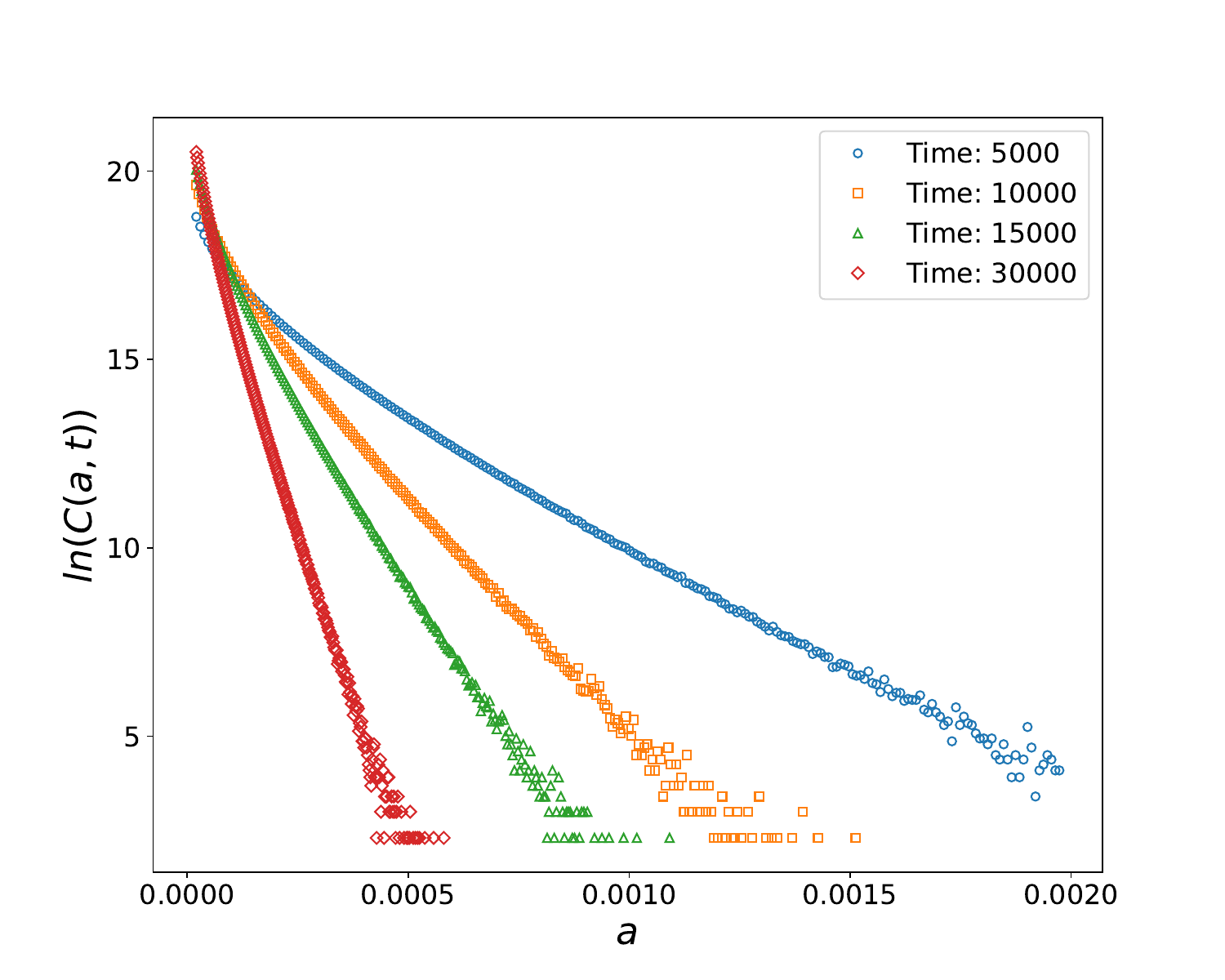}
\label{fig:3a}
}
\subfloat[]
{
\includegraphics[height=4.0 cm, width=4.0 cm, clip=true]
{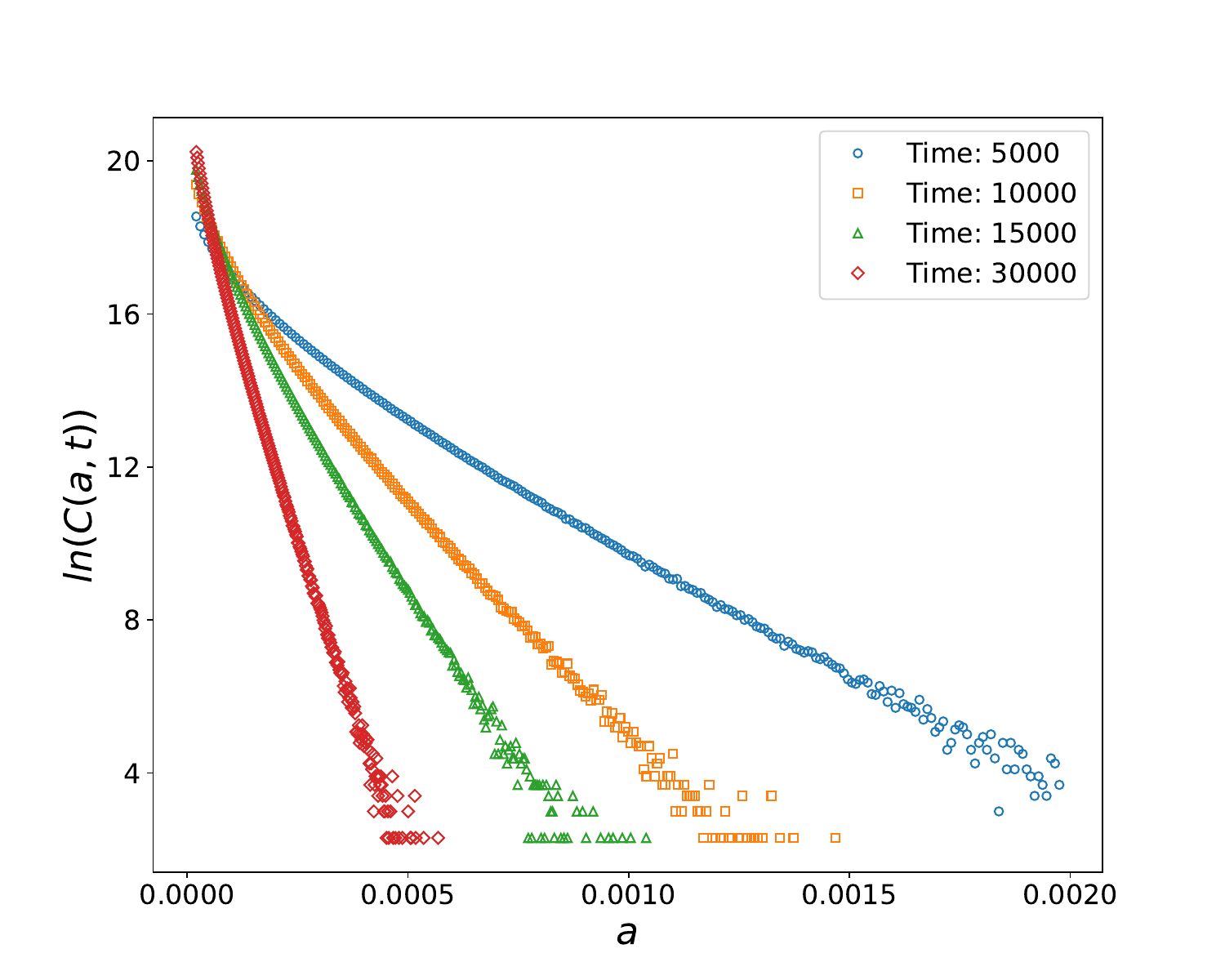}
\label{fig:3b}
}

\subfloat[]
{
\includegraphics[height=4.0 cm, width=4.0 cm, clip=true]
{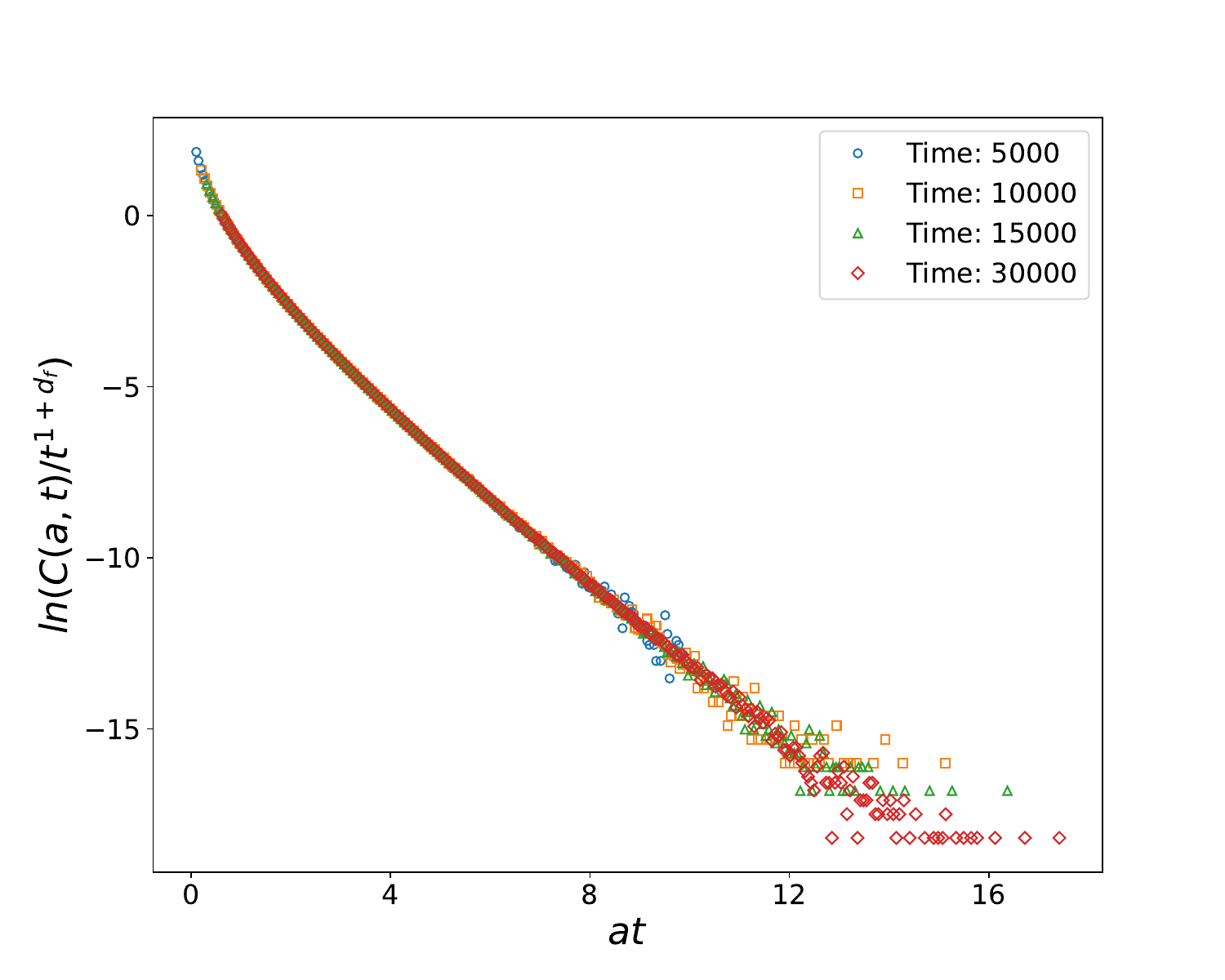}
\label{fig:3c}
}
\subfloat[]
{
\includegraphics[height=4.0 cm, width=4.0 cm, clip=true]
{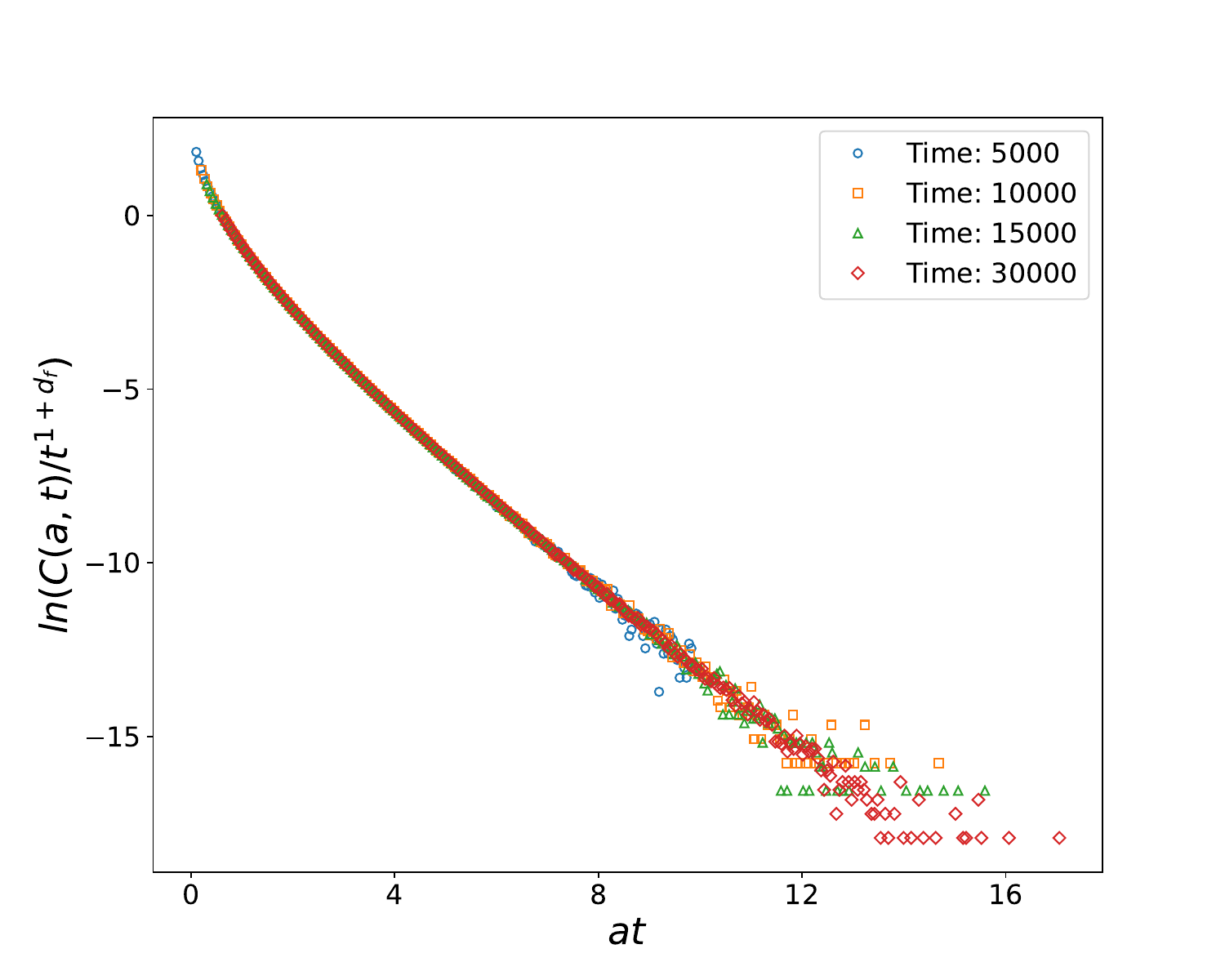}
\label{fig:3d}
}
\caption{The natural logarithm of the area distribution function $C(a,t)$ plotted as a function of $a$ for $t = 5000,\ 10{,}000,\ 15{,}000,$ and $30{,}000$: (a) $q = 0.95$ and (b) $q = 0.85$. In both cases, the plots are approximately linear in the tail region, indicating an exponential form of the distribution. The same data as in Figs. (\ref{fig:3a}) and (\ref{fig:3b})  are used to plot the dimensionless quantities $\ln\big(C(a,t),t^{-\sqrt{3+q}}\big)$ versus $a t$ in (c) for $q=0.95$ and in (d) for $q=0.85$. An excellent collapse of the distinct curves shown in Fig. (\ref{fig:3}) is observed.
} 
\label{fig:3}
\end{figure}

We characterize each block solely by its area, neglecting geometric details and thus reducing the problem to an effective one-dimensional size distribution. We define $C(a,t)\,da$ as the number of blocks with areas in the interval $[a,a+da]$ at time $t$, obtained numerically by binning areas with width $\delta a$ and normalizing by $\delta a$.
The total area of surviving blocks decays as
\begin{equation}
\label{dc_1}
\int_0^\infty a\,C(a,t)\,da =M(2,2;t) \sim t^{-(2-\sqrt{3+q})},
\end{equation}
while the total number of blocks grows as
\begin{equation}
\label{eq:n}
\int_0^\infty C(a,t)\,da =M(1,1;t)\sim t^{(\sqrt{3+q}-1)}.
\end{equation}
These correspond to the moments $M(2,2;t)$ and $M(1,1;t)$, respectively, providing a direct link between microscopic block statistics and macroscopic scaling behavior.
Together, these quantities show how the evolving block-area distribution simultaneously captures both the depletion of total mass and the proliferation of blocks during the stochastic growth process.

To test whether the block-area distribution $C(a,t)$ exhibits dynamic scaling, we employ dimensional analysis via the Buckingham $\Pi$ theorem \cite{barenblatt1996scaling}. From Eq.~(\ref{eq:mnt}), the mean area of surviving blocks scales as $\langle a(t)\rangle \sim t^{-1}$, implying that time sets the natural scale for area. Hence, $a$ and $t$ are not independent, and the dependence of $C(a,t)$ reduces to a single dimensionless variable
\begin{equation}
\label{xi}
\xi = a t .
\end{equation}
Consistency then requires $C(a,t)$ to scale as $C(a,t)\sim t^{\theta}$, allowing the definition of a dimensionless distribution
\begin{equation}
\label{Pi}
\Pi = \frac{C(a,t)}{t^{\theta}} .
\end{equation}
This establishes dynamic scaling, whereby $C(a,t)$ collapses onto a universal function of the single scaling variable $\xi=at$.

By definition, a dimensionless quantity remains invariant under a rescaling of the underlying dimensional variables. Accordingly, the numerical value of $\Pi$ must remain unchanged if time $t$ is rescaled by an arbitrary factor. Nevertheless, $\Pi$ may still depend on the dimensionless governing parameter $\xi$, and we may therefore write
\begin{equation}
\label{Pi_sim}
\Pi \sim \phi(\xi),
\end{equation}
where $\phi(\xi)$ is a dimensionless function. Substituting the definition of $\Pi$ from Eq.~(\ref{Pi}) into Eq.~(\ref{Pi_sim}), we obtain the dynamic scaling form of the distribution function,
\begin{equation}
\label{disfunc}
C(a,t) \sim t^{\theta}\phi(at),
\end{equation}
with $\phi(\xi)$ denoting the scaling function and $\theta$ an a priori unknown scaling exponent.

The exponent $\theta$ is determined by the fractal dimension $d_f$ of the lattice, which is, in fact, the dimension of the underlying skeleton on which the measures $\{p_i\}$ are distributed. The number of blocks grows algebraically following Eq. (\ref{eq:n}).
Introducing the yardstick $\delta \sim t^{-1/2}$, Eq. (\ref{eq:n}) can be rewritten in the standard fractal form
\begin{equation}
N(\delta) \sim \delta^{-d_f},
\end{equation}
from which we identify
\begin{equation}
d_f = \sqrt{3+q}-1.
\end{equation}
The $d_f$-th moment of the distribution is a conserved quantity, and this conservation uniquely fixes the value of the scaling exponent $\theta$ \cite{hassan2002randomness,hassan2014dyadic}. Enforcing this constraint yields
\begin{equation}
\theta = 1 + d_f .
\end{equation}
Consequently, the block-area distribution obeys the dynamic scaling form
\begin{equation}
\label{Ct}
C(a,t) \sim t^{\sqrt{3+q}} \phi(at).
\end{equation}
This result has a clear and testable implication. Plots of $C(a,t)$ versus $a$ at different times $t$ will, as shown in Figs. (\ref{fig:3a}) and (\ref{fig:3b}) for $q=0.95$ and for $q=0.85$. 

We next probe the explicit $a$-dependence of the area distribution $C(a,t)$. 
Plotting $\ln\!\left(C(a,t)t^{-\sqrt{3+q}}\right)$ against $at$ (Figs.~(\ref{fig:3c}), \ref{fig:3d}) collapses all curves from Figs.~(\ref{fig:3a}) and (\ref{fig:3b}) onto a single master curve $\phi(\xi)$, providing compelling evidence for dynamic scaling and temporal self-similarity. 
The collapsed curves are linear in the tail, with slopes equal to $t$, implying
\begin{equation}
\phi(a)\sim e^{-at},
\label{scaling}
\end{equation}
 yielding
\begin{equation}
C(a,t)\sim t^{\sqrt{3+q}}e^{-at}.
\label{area_final}
\end{equation}
It demonstrates that while dimensional quantities vary with time, the dimensionless distribution remains invariant. Hence, $C(a,t)$ obeys dynamic scaling, and lattice snapshots at different times are self-similar.

\section{Scale free properties WPSPL}

\begin{figure}
\centering
\subfloat[]
{
\includegraphics[height=4.0 cm, width=4.0 cm, clip=true]
{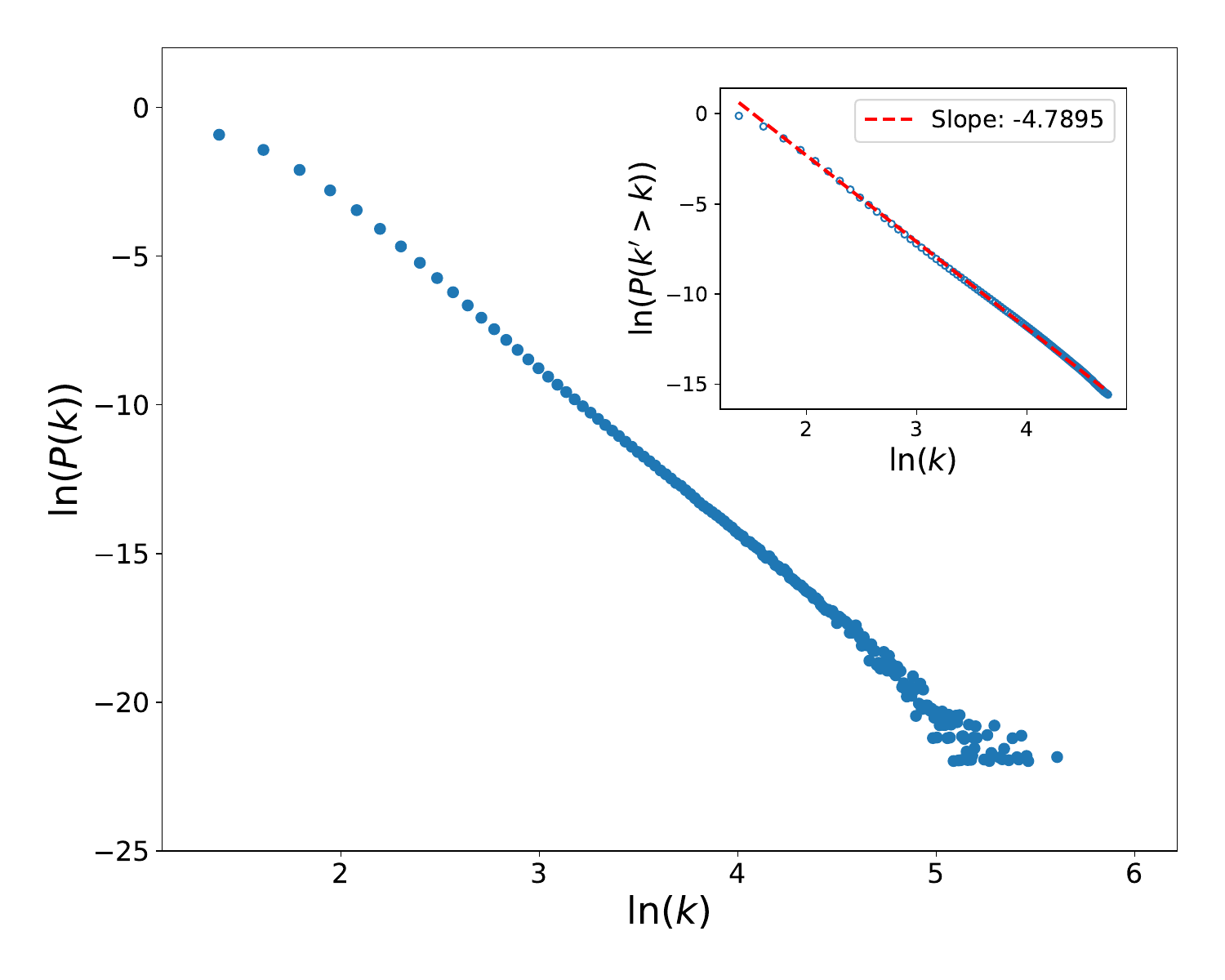}
\label{fig:4a}
}
\subfloat[]
{
\includegraphics[height=4.0 cm, width=4.0 cm, clip=true]
{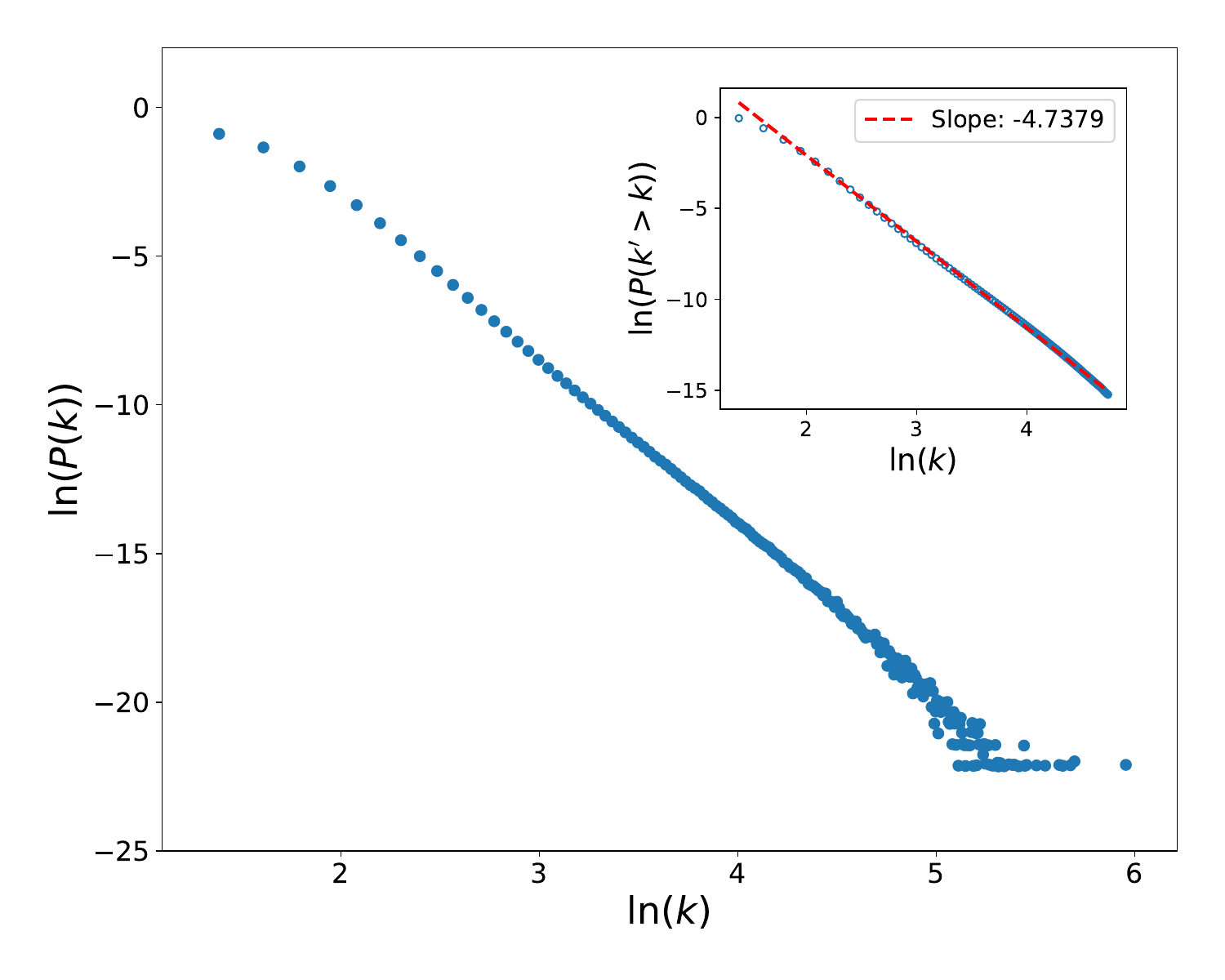}
\label{fig:4b}
}
\caption{The degree distribution $P(k)$ for the dual of the WPSPL network is shown on a log--log scale in (a) for $q=0.85$ and in (b) for $q=0.95$. The data points are averages over $50000$ independent realizations. In both cases, the distributions are approximately linear, indicating a power-law behavior. However, the heavy tail implies poor statistics for large $k$, making a direct estimation of the exponent from $P(k)$ unreliable. To mitigate this problem, the insets display the cumulative degree distribution $P(k' > k)$ computed from the same data as it remove the fat tail and provides a more reliable estimate of the scaling exponent. The exponent of $P(k)$ is then obtained by adding one to the exponent measured from the cumulative distribution.
} 
\label{fig:4}
\end{figure}

Here we focus on the coordination number distribution of the blocks in the WPSPL. 
In a regular square lattice, the coordination number is constant and equal to $4$. 
In contrast, the coordination number in the WPSPL is neither constant nor can be characterized by a typical mean value; rather, it is a random quantity that evolves with time. Thus, the coordination-number disorder in the WPSPL is of {\it annealed} type.
If the center of each block is regarded as a node and the common border between two neighboring blocks as a link, the resulting structure forms the dual network of the WPSPL (DWPSPL). 
In this representation, the coordination number of a block corresponds to the degree of the associated node. Defining each step of the growth algorithm as one unit of time and imposing periodic boundary conditions, we measure the fraction of blocks having $k$ nearest neighbors,
$\frac{N_k(t)}{N(t)}$, where $N_k(t)$ is the number of blocks with coordination number $k$ and $N(t)$ is the total number of blocks having any number of coordination number. In the network representation, $P(k)=\frac{N_k(t)}{N(t)}$ therefore corresponds to the degree distribution of the DWPSPL.

In Fig.(\ref{fig:4}) we plot $\ln P(k)$ versus $\ln k$ using data averaged over $50000$ independent realizations. 
The resulting curve is approximately linear, indicating that the degree distribution follows a power law
\begin{equation}
\label{degreedistribution}
P(k) \sim k^{-\gamma}.
\end{equation}
However, the distribution exhibits a heavy (fat) tail, a characteristic feature of scale-free networks, corresponding to highly connected {\it hub} nodes. 
Such sparse statistics in the tail region make a direct estimation of the exponent $\gamma$ unreliable. 
To reduce this noise, we also compute the cumulative degree distribution $P(k' > k)$ \cite{AlbertBarabasi2002}. The inset of Fig. (\ref{fig:4}) shows a plot of $\ln P(k' > k)$ versus $\ln k$ using the same data. In this representation, the tail fluctuations are naturally smoothed, allowing a more reliable estimation of the scaling exponent.

A linear fit to the plot of $\ln P(k' > k)$ in the inset yields a slope of $\gamma-1 = 4.7895$ for $q=0.85$ and $\gamma-1 = 4.7379$ for $q=0.95$. This implies that the degree distribution in Fig.~\ref{fig:4} follows a power law with exponents $\gamma = 5.7895$ and $\gamma = 5.7379$ for $q=0.85$ and $q=0.95$, respectively. We further observe that as $q$ increases, the exponent decreases to $\gamma = 5.66$ when $q=1$ \cite{ref.hassan_njp, ref.hassan_jpc}.  Barabási has argued that power-law degree distributions arise from preferential attachment, often described as a ``rich-get-richer'' mechanism \cite{Achlioptas2009,ref.barabasi_decade_0}. The growth dynamics of the WPSPL contains a similar ingredient. In this model, a node gains new links only when one of its neighboring blocks is selected for subdivision. Consequently, nodes with larger coordination numbers have a higher probability of acquiring additional links. This effectively leads to preferential growth in degree. Employing this indirect preferential attachment rule, in 2016 we proposed the mediation-driven attachment network. We showed that the resulting network is indeed scale-free, with a spectrum of exponents depending on the number of links with which incoming nodes join the network \cite{ref.hassan_liana, HassanSarker2020}.

\section{Random bond percolation}

To implement bond percolation on the WPSPL, we first define its dual representation. 
Following the construction of a regular square lattice, we associate a site with the center of each surviving block and a bond with the shared boundary between neighboring blocks. 
This yields a dual network of sites connected by bonds, on which bond percolation is naturally defined. Unlike regular lattices, however, each site carries a weight equal to the area of the block it represents. Consequently, clusters formed by occupied bonds are characterized not by the number of sites but by their total area, given by the sum of the areas of all sites belonging to the cluster.

Once the dual of the WPSPL has been constructed, each site is assigned a unique label,
$i = 1, 2, 3, \ldots$, in order to keep track of the endpoints of every bond and to unambiguously identify the pair of sites connected by each bond. After completing this labeling procedure, all bonds are initially removed so that the system starts from a configuration of $N$ isolated sites, each carrying an associated area corresponding to the block it represents. Bond occupation is then implemented following the Newman–Ziff (N-Z) algorithm~\cite{NewmanZiff2000,NewmanZiff2001}. Specifically, all bonds are first randomized and subsequently reoccupied one by one. After each bond addition, clusters are dynamically updated, and the cluster size is defined not by the number of sites but by the sum of the areas associated with the sites belonging to the cluster.

Although the WPSPL is statistically self-similar, it is intrinsically disordered; consequently, the total number of bonds at a given time $t$ varies between realizations. 
Hence, the bond count must be determined separately for each dual lattice. 
All observables are obtained by averaging over many independent realizations. 
For efficiency, we perform multiple percolation runs—1000 independent Newman-Ziff (NZ) runs—on each realization at fixed $t$, and then repeat the procedure for 10000 independent realizations. 
The NZ algorithm yields observables $X_n$ directly as functions of the number of occupied bonds (or sites) $n$. The resulting data are then used in the convolution relation
\begin{equation}
\label{eq:convolution}
X(p)=\sum_{n=1}^{N} p^{n}(1-p)^{N-n} X_n,
\end{equation}
which allows $X(p)$ to be obtained for any occupation probability $p$. The binomial weight factor associated with each $n$ at a given $p$ is automatically incorporated through the convolution \cite{ref.Ziff}, yielding smooth and well-resolved curves for $X(p)$.

\subsection{Spanning probability $W(p)$}

To compute the spanning probability, we perform $\Omega$ independent percolation realizations. 
In each realization, bonds are occupied sequentially and the occupation number $n_c^{(i)}$ at which a spanning cluster first appears is recorded. 
The spanning probability is then defined as
\begin{equation}
W(n,N)=\frac{1}{\Omega}\sum_{i=1}^{\Omega}\Theta\!\left(n-n_c^{(i)}\right),
\end{equation}
where $n$ is the number of occupied bonds, $n_c^{(i)}$ denotes the spanning threshold of the $i$-th realization, and $\Theta(x)$ is the Heaviside step function. 
Thus, $W(n,N)$ represents the cumulative distribution of the realization-dependent spanning threshold $n_c$.

\begin{figure}
\centering
\subfloat[]
{
\includegraphics[height=4.2 cm, width=4.2 cm, clip=true]
{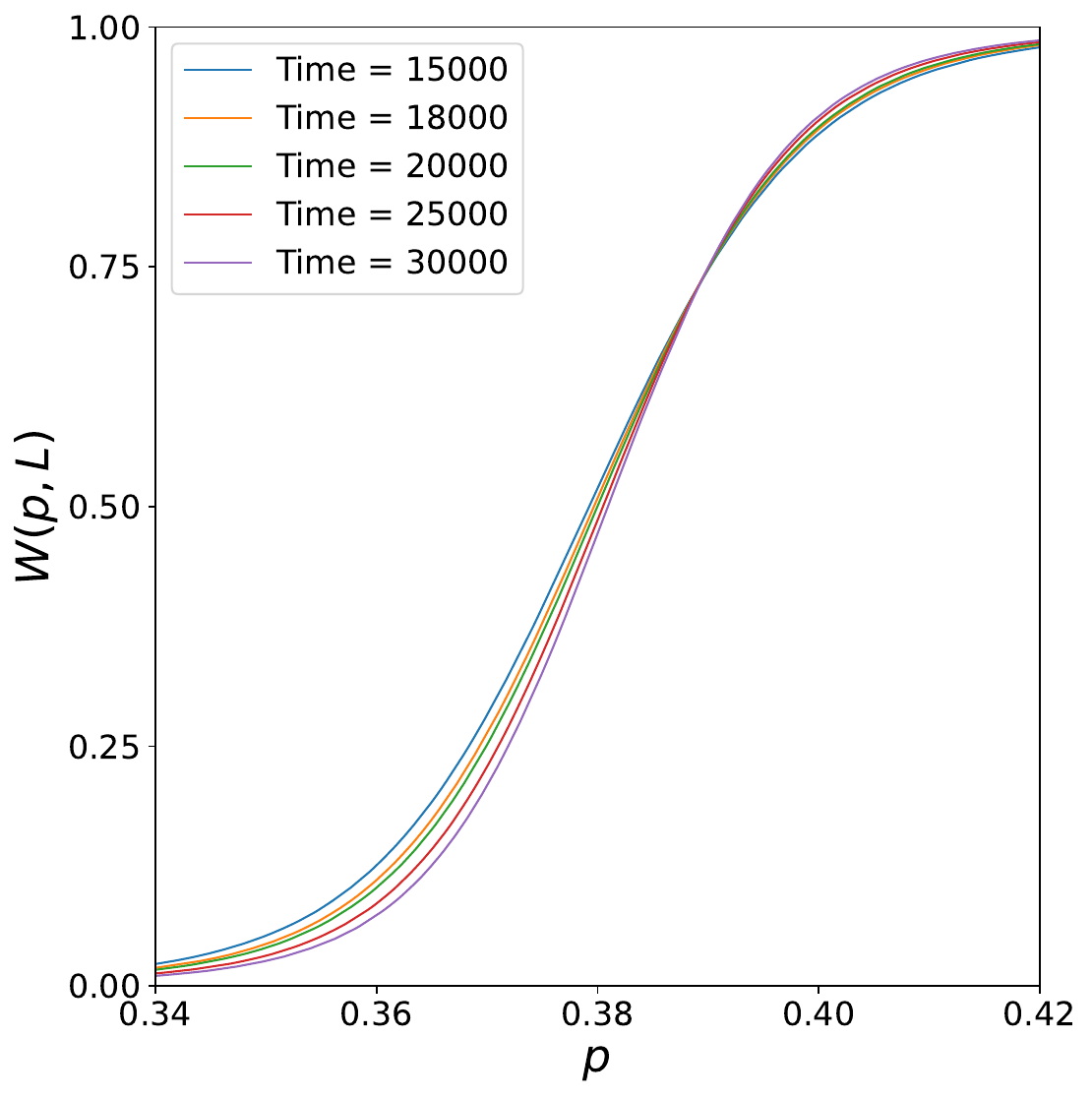}
\label{fig:spanning_1}
}
\subfloat[]
{
\includegraphics[height=4.2 cm, width=4.2 cm, clip=true]
{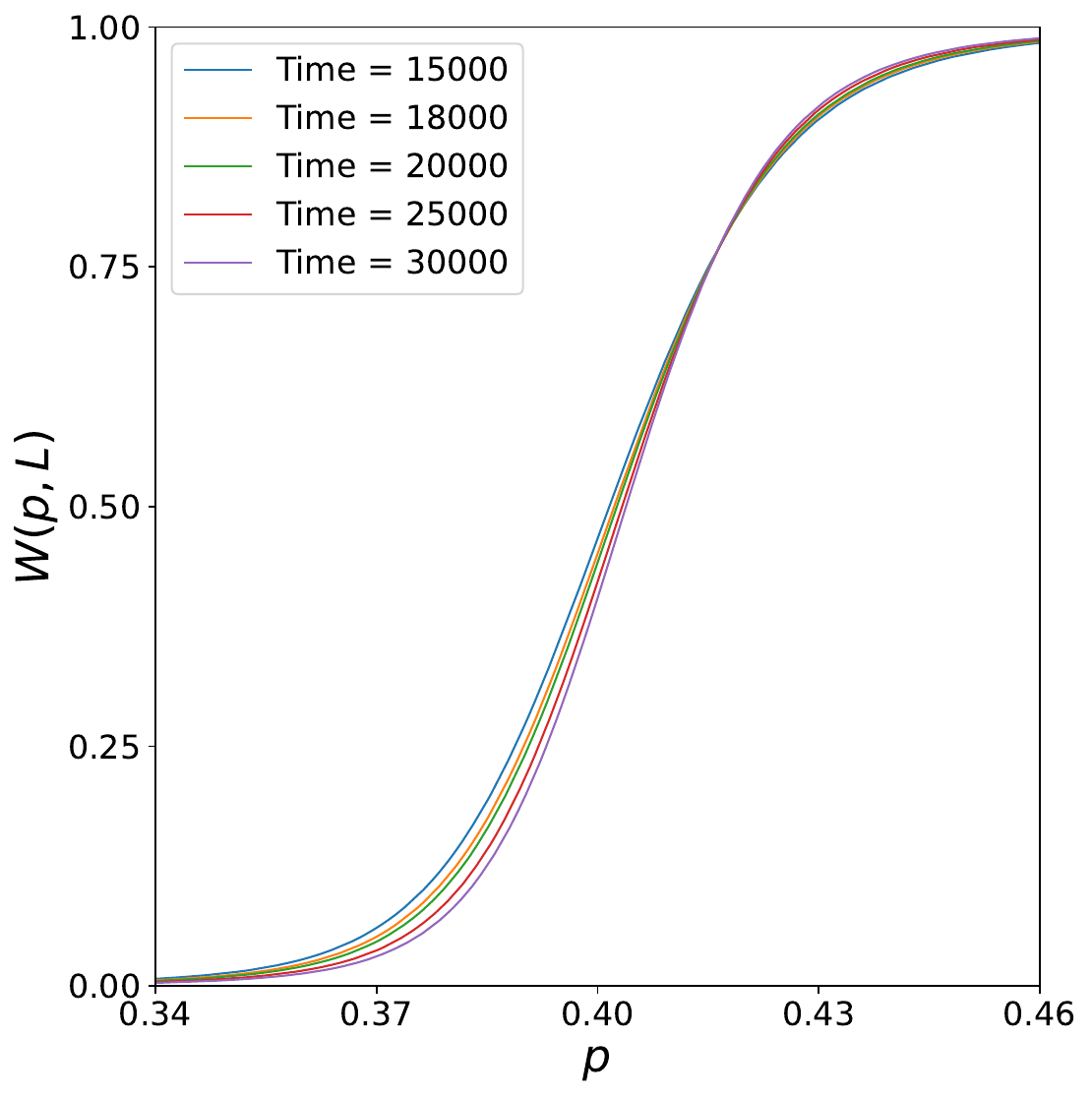}
\label{fig:spanning_2}
}

\subfloat[]
{
\includegraphics[height=4.2 cm, width=4.2 cm, clip=true]
{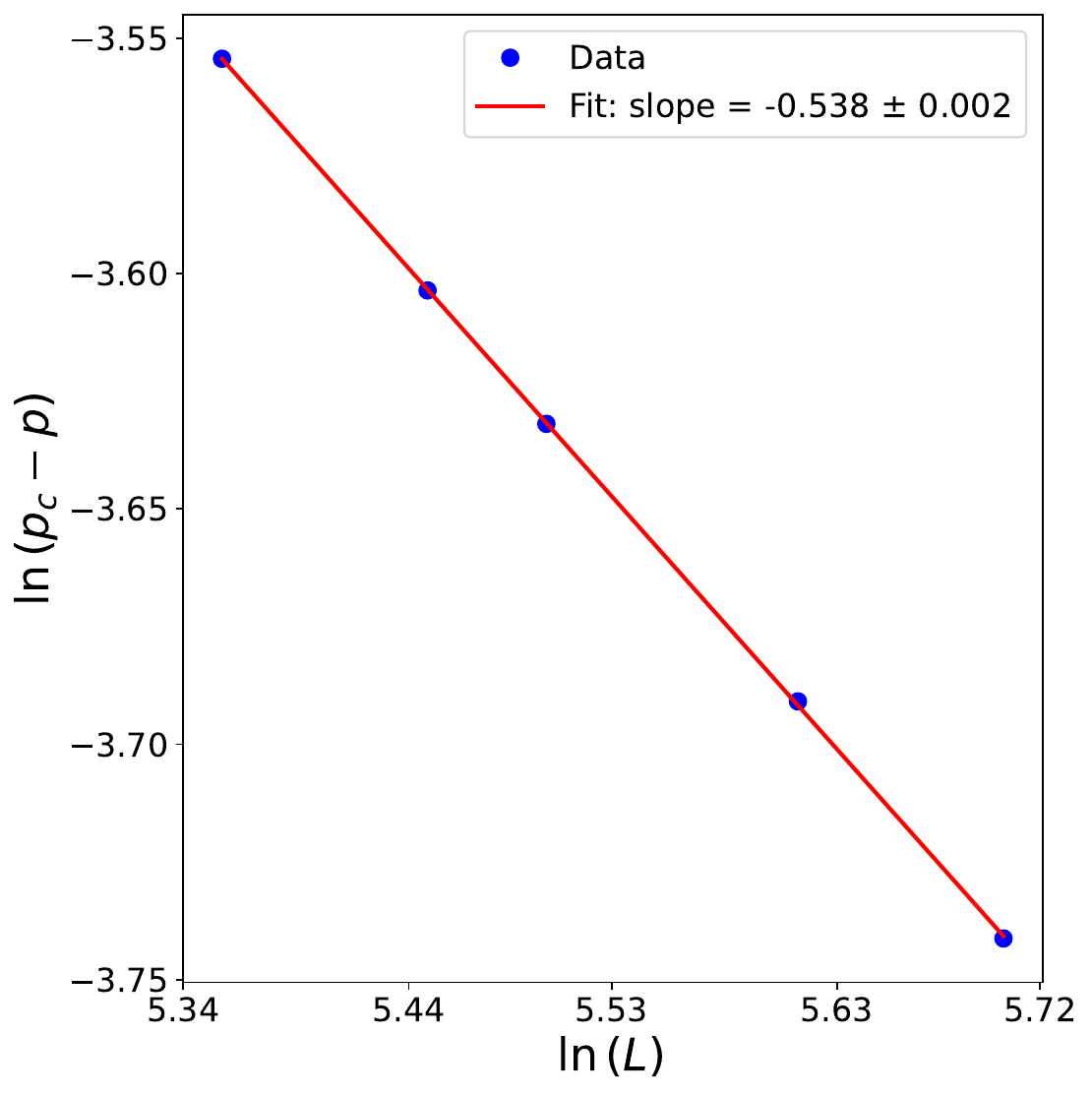}
\label{fig:spanning_3}
}
\subfloat[]
{
\includegraphics[height=4.2 cm, width=4.2 cm, clip=true]
{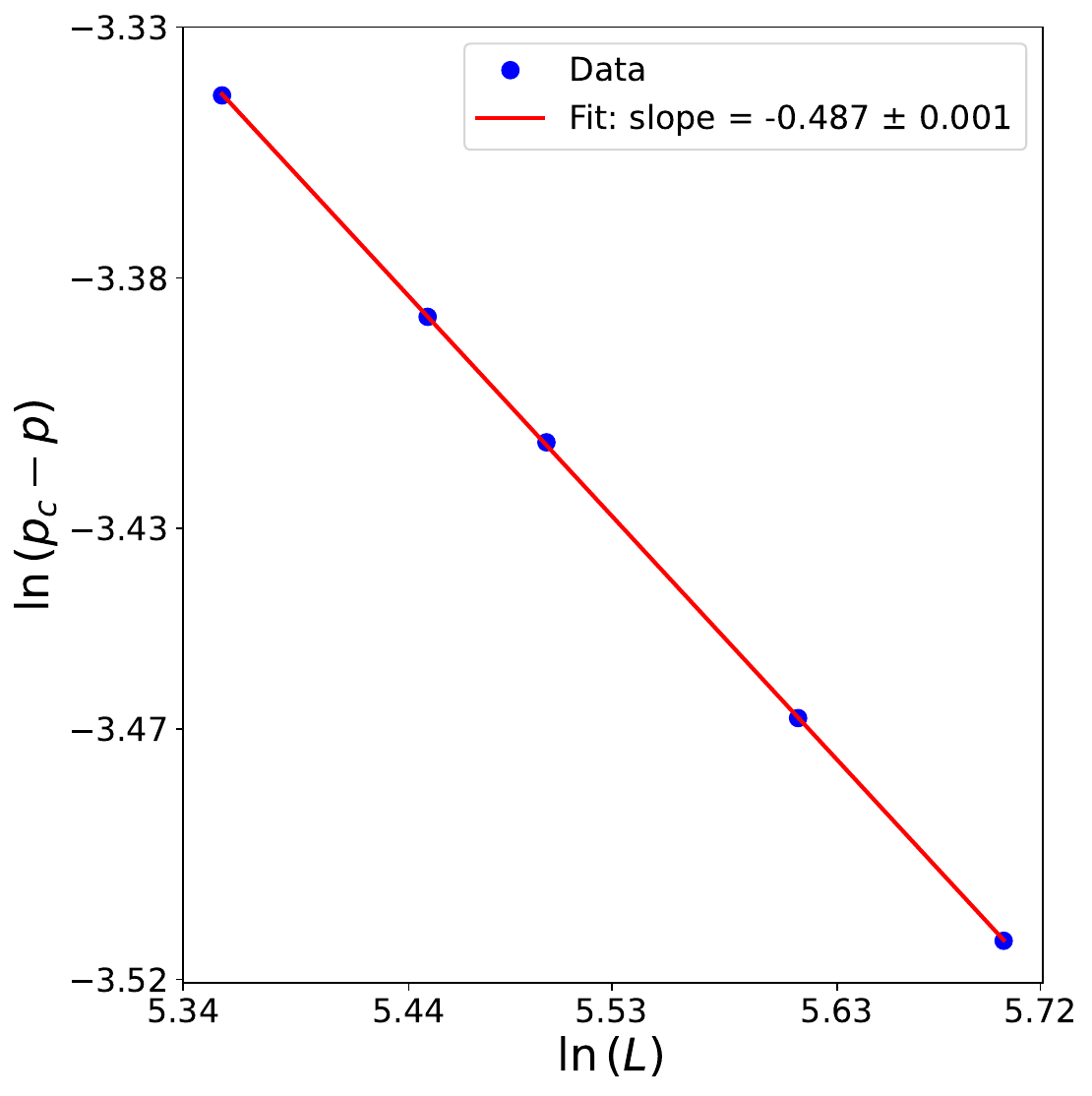}
\label{fig:spanning_4}
}
\caption{Spanning probability $W(p,L)$ as a function of $p$ for bond percolation on the WPSPL is shown for different system sizes at block-retaining probabilities (a) $q = 0.90$ and (b) $q = 0.85$. From the simulations, the estimated percolation thresholds are $p_c = 0.389389$ for $q = 0.90$ and $p_c = 0.416496$ for $q = 0.85$. Panels (c) and (d) show plots of $\log(p_c - p)$ versus $\log L$ for $q = 0.90$ and $q = 0.85$, respectively. The resulting slopes yield the correlation-length exponents $1/\nu = 0.538 \pm 0.002$ for $q = 0.90$ and $1/\nu = 0.487 \pm 0.001$ for $q = 0.85$.
} 
\label{fig:spanning}
\end{figure}

Once $W(n,N)$ is computed, it can be substituted into Eq.~(\ref{eq:convolution}) to obtain the spanning probability $W(p,N)$. 
Numerical simulations for different system sizes yield curves such as those in Fig.~\ref{fig:spanning}, which intersect near a common point. 
This intersection provides a robust estimate of the critical occupation probability $p_c$: for finite $N$, $W_{\rm span}(p)$ is analytic and can be expanded as a polynomial, so exact intersections do not exist. 
However, the approximate intersection converges to $p_c$ in the thermodynamic limit, where finite-size corrections are negligible. 
Thus, the intersection not only reflects geometry but also signals the percolation transition and its universal critical behavior.

\begin{figure}
\centering
\subfloat[]
{
\includegraphics[height=4.2 cm, width=4.2 cm, clip=true]
{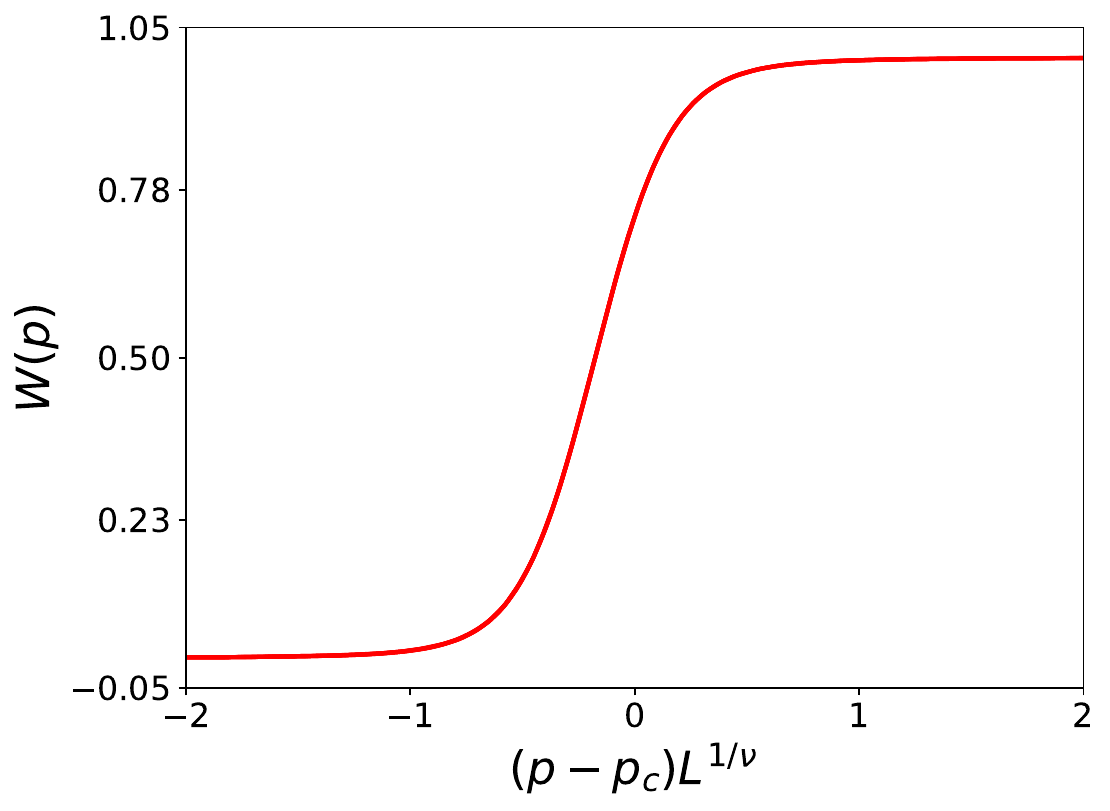}
\label{fig:spanning_collapsed_1}
}
\subfloat[]
{
\includegraphics[height=4.2 cm, width=4.2 cm, clip=true]
{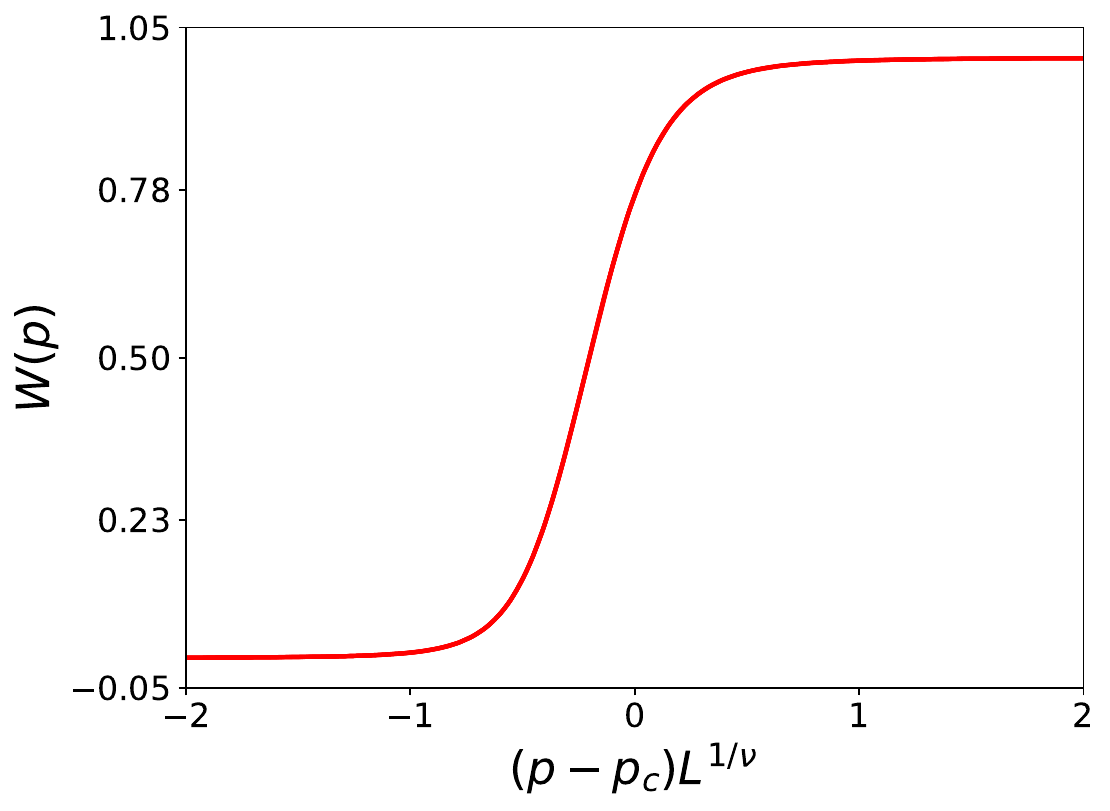}
\label{fig:spanning_collapsed_2}
}
\caption{We show plots of $W(p)$ versus $(p_c - p) L^{1/\nu}$ for $q = 0.90$ and $q = 0.85$ in panels (a) and (b), respectively. We observe that all the distinct curves from Figs.~\ref{fig:spanning_1} and \ref{fig:spanning_2} collapse onto their respective universal scaling curves.} 
\label{fig:spanning_collapsed}
\end{figure}

A key feature of the $W(p)$ versus $p$ curves is their systematic shift toward the critical point $p_c$ as the system size $L$ increases. On both sides of $p_c$, all points move closer, with the shift sharper for $p<p_c$ and small $W(p,L)$. This finite-size shift can be quantified by drawing a horizontal line at a fixed $W$, where the curve separation is maximal, and measuring the distance $p_c - p$ for each $L$. A log–log plot of $p_c - p$ versus $L$ yields  
\begin{equation}
\label{eq:sp1}
p_c - p \sim L^{-1/\nu}.
\end{equation}
From Figs.~(\ref{fig:spanning_3}) and (\ref{fig:spanning_4}), we estimate $1/\nu = 0.538 \pm 0.002$ for $q = 0.90$ and $1/\nu = 0.487 \pm 0.001$ for $q = 0.85$. Equation~(\ref{eq:sp1}) implies that in the thermodynamic limit $L \to \infty$, all finite-size estimates converge to $p_c$, and the spanning probability approaches a step function:
\begin{equation}
\lim_{L \rightarrow \infty} W(p,L) =
\begin{cases}
0, & p < p_c, \\
1, & p \ge p_c.
\end{cases}
\end{equation}
Equation~(\ref{eq:sp1}) also motivates a dimensionless scaling variable $(p - p_c)L^{1/\nu}$. When plotted as $W(p)$ versus $(p_c - p)L^{1/\nu}$, all distinct curves collapse onto a universal curve, as shown in Figs.~(\ref{fig:spanning_collapsed_1}) and (\ref{fig:spanning_collapsed_2}) for $q=0.90$ and $q=0.85$, respectively.

\subsection{Analog of Thermodynamic Response Functions in percolation}

Specific heat and susceptibility are two fundamental thermodynamic response functions, quantifying how a system reacts to changes in its external control parameters. They are defined as derivatives of suitable thermodynamic quantities and play a central role in the study of phase transitions and critical phenomena \cite{stanleyBook}. In thermodynamics, at fixed external parameters, specific heat is defined as
\begin{equation}
\label{eq:sp}
C = T \left( \frac{\partial S}{\partial T} \right),
\end{equation}
which quantifies thermal fluctuations and is purely thermodynamic in origin.
The susceptibility, on the other hand, measures the response of an order parameter to its conjugate external field $h$, which tends to enhance the ordering of the system. It is defined as
\begin{equation}
\label{eq:sus}
\chi = \left( \frac{\partial \langle O \rangle}{\partial h} \right)_T,
\end{equation}
where $\langle O \rangle$ denotes the order parameter. Specific heat probes the thermal response of the system, while susceptibility probes the response of an order parameter to its conjugate field. Both are thermodynamic response functions that encode essential information about the collective behavior of many-body systems.

In percolation theory, a key challenge is constructing meaningful analogs of thermodynamic response functions—specifically, the specific heat and susceptibility—within a non-thermal framework. In equilibrium systems, the specific heat measures how disorder (entropy) responds to temperature changes, with higher temperature typically reducing order. Translating this to percolation requires a well-defined notion of entropy for random connectivity and the identification of a control parameter that mimics temperature in the absence of thermal fluctuations. Once established, one can study the evolution of disorder near the percolation threshold and its critical behavior. Similarly, an analog of susceptibility must be defined to quantify the response of the percolation order parameter to an external field, completing the thermodynamic analogy and enabling a unified scaling description of percolation.

\begin{figure}
\centering
\subfloat[]
{
\includegraphics[height=4.2 cm, width=4.2 cm, clip=true]
{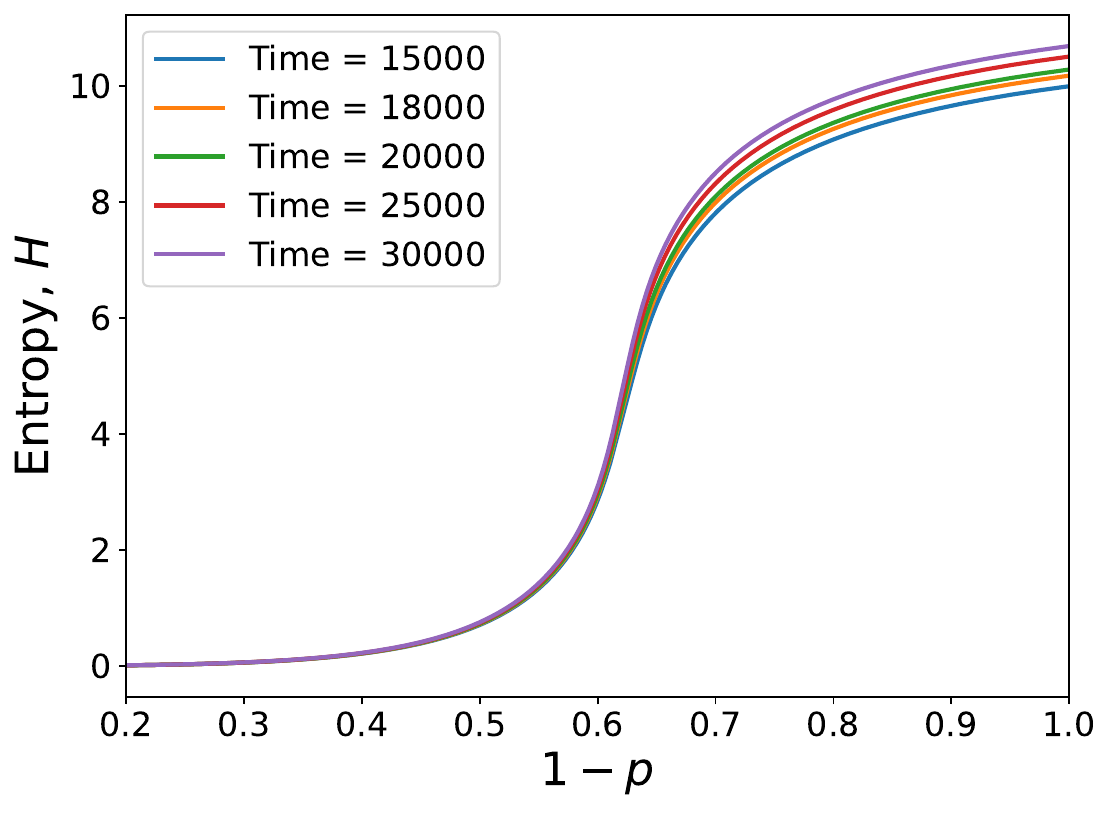}
\label{fig:entropy_1}
}
\subfloat[]
{
\includegraphics[height=4.2 cm, width=4.2 cm, clip=true]
{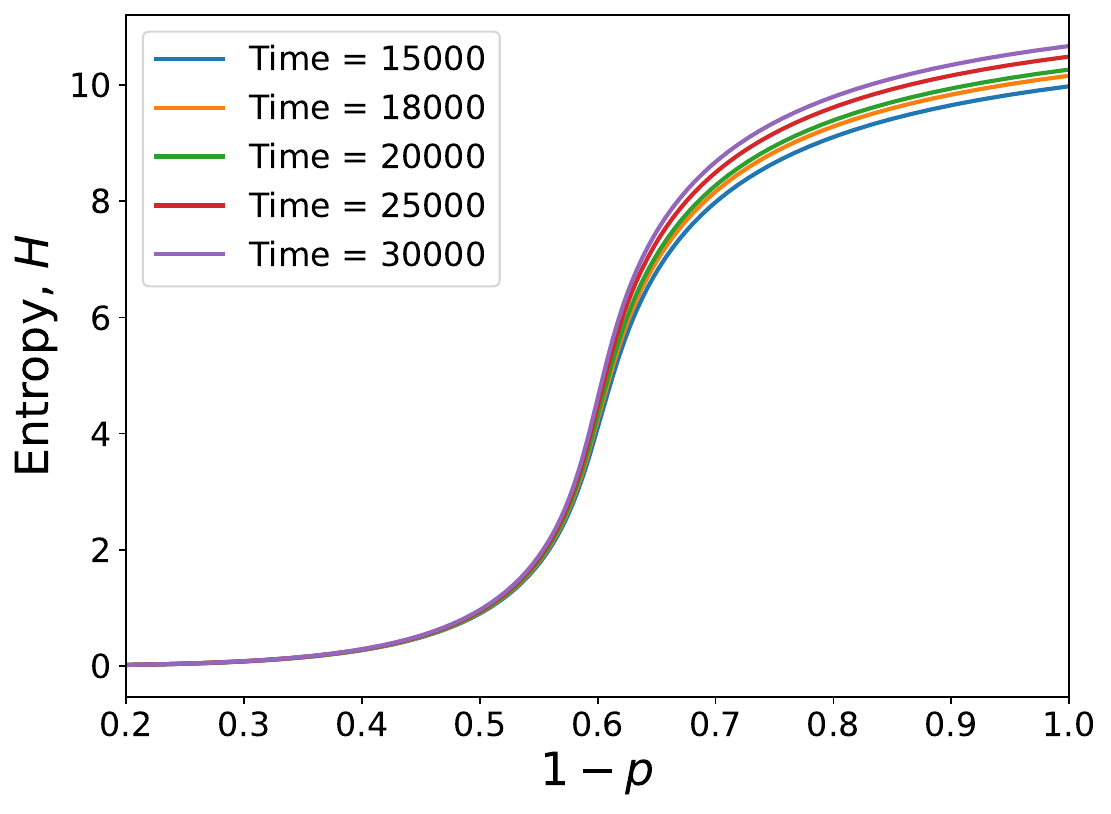}
\label{fig:entropy_2}
}
\caption{Entropy $H$ plotted as a function of $(1-p)$ for different system sizes and $q$ values: (a) $q=0.90$ and (b) $q=0.85$.}
\label{fig:entropy}
\end{figure}

\subsection{Entropy and order parameter}

We consider random bond percolation, where each site initially forms an isolated cluster of unit size. As bonds are occupied, clusters merge and grow. At an arbitrary stage, let there be $k$ distinct clusters with sizes $s_1, s_2, \dots, s_k$. Within a microcanonical description, the number of configurations consistent with this cluster structure is
\begin{equation}
\Omega_k = \frac{N!}{s_1! , s_2! \cdots s_k!},
\end{equation}
yielding the Boltzmann entropy
\begin{equation}
S_k = \ln \Omega_k,
\end{equation}
where $k_B=1$. Using Stirling’s approximation, the entropy per site reduces to
\begin{equation}
\label{eq:shannon}
\frac{S_k}{N} = -\sum_{i=1}^k \mu_i \ln \mu_i,
\end{equation}
with $\mu_i = s_i/N$ denoting the cluster picking probability. Eq. (\ref{eq:shannon}) is formally identical to the Shannon entropy and provides a normalized measure of disorder per site in percolation.
Fig. (\ref{fig:entropy}) shows $H$ as a function of $1-p$ for different system sizes. As in thermal phase transitions, the entropy exhibits a sigmoidal crossover from a low-entropy ordered phase to a high-entropy disordered phase. This close correspondence identifies $1-p$ as the natural analog of temperature in percolation, consistent with its non-negativity and monotonic control of disorder.

\begin{figure}
\centering
\subfloat[]
{
\includegraphics[height=4.2 cm, width=4.2 cm, clip=true]
{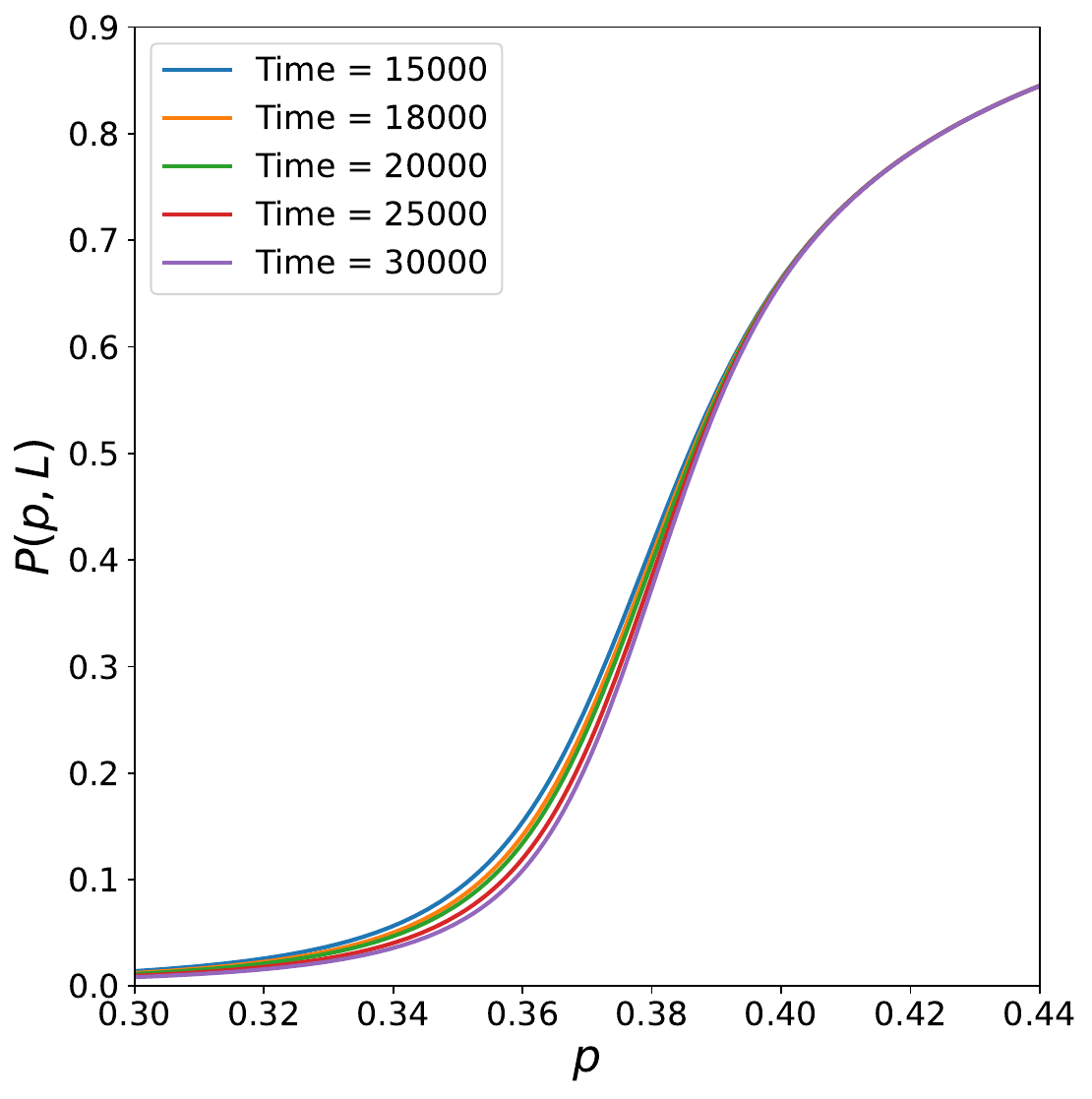}
\label{fig:op_1}
}
\subfloat[]
{
\includegraphics[height=4.2 cm, width=4.2 cm, clip=true]
{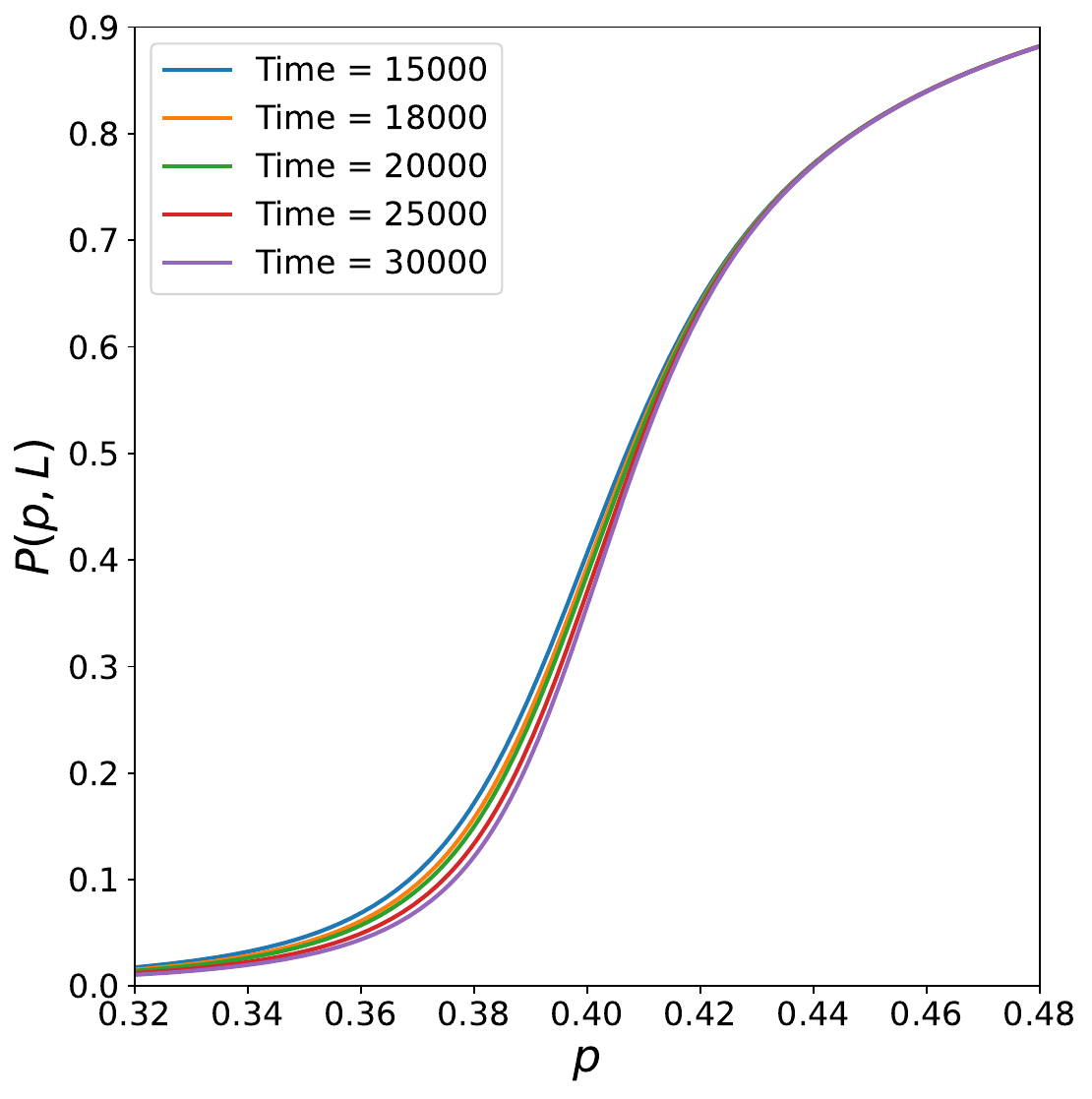}
\label{fig:op_2}
}

\subfloat[]
{
\includegraphics[height=4.2 cm, width=4.2 cm, clip=true]
{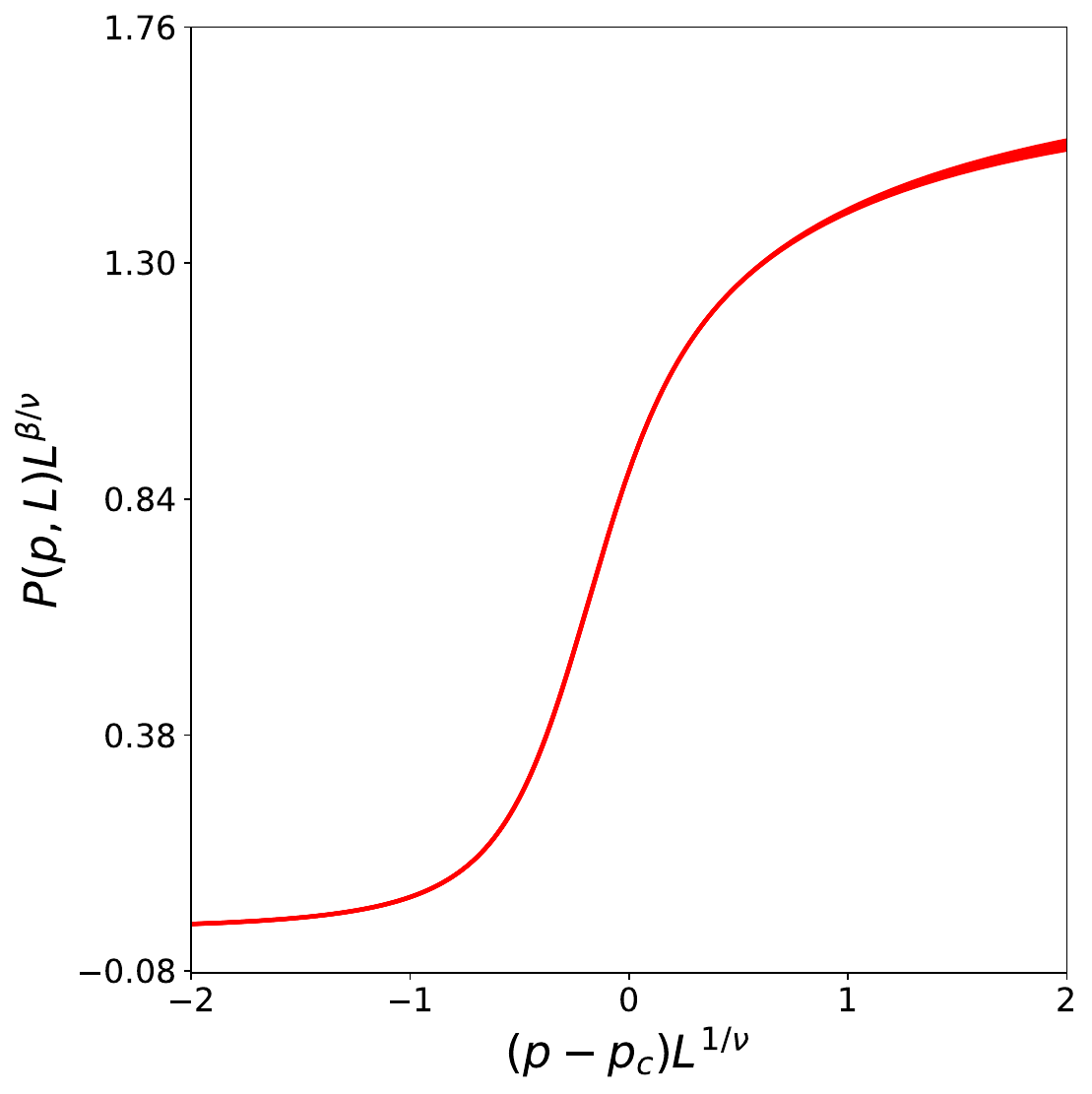}
\label{fig:op_1_collapsed}
}
\subfloat[]
{
\includegraphics[height=4.2 cm, width=4.2 cm, clip=true]
{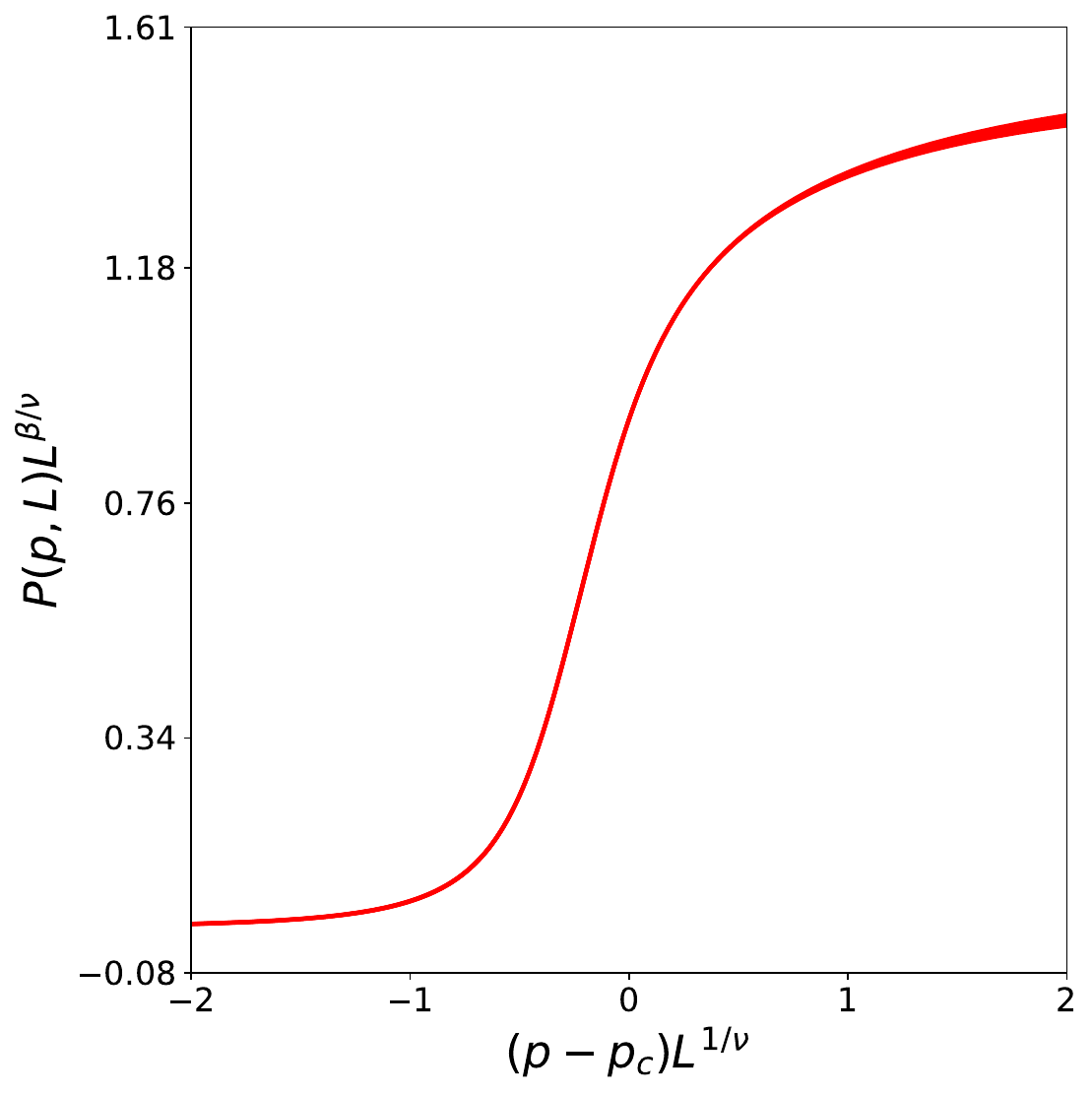}
\label{fig:op_2_collapsed}
}
\caption{Order parameter $P(p,L)$ plotted as a function of $p$ for (a) $q=0.90$ and (b) $q=0.85$. Plots of $P(p,L)L^{-{\beta}/{\nu}}$ versus $(p_c - p) L^{1/\nu}$ for $q = 0.90$ and $q = 0.85$ are shown in panels (c) and (d), respectively. We observe that all the distinct curves from (a) and (b) collapse onto their respective universal scaling curves.} 
\label{fig:op}
\end{figure}

The deep connection between percolation and thermal phase transitions was established by Kasteleyn and Fortuin through their mapping of percolation onto the $q$-state Potts model, in which bond percolation is recovered in the singular limit $q \to 1$ \cite{KF1969,FK1972}. Within this framework, the relative size of the largest cluster,
\begin{equation}
P(p,L) = \frac{S_{\rm max}}{N},
\end{equation}
serves as an order parameter analogous to magnetization in magnetic systems. For $p < p_c$, all clusters remain finite and $P \to 0$ in the thermodynamic limit. For $p > p_c$, a macroscopic (giant) cluster emerges, and the order parameter grows continuously as
\begin{equation}
P \sim (p-p_c)^{\beta},
\end{equation}
where $\beta$ is the associated critical exponent. This continuous onset closely parallels the paramagnetic–ferromagnetic transition, firmly establishing percolation as a geometric realization of a second-order phase transition. In conventional magnetic systems, an external magnetic field $h$ enhances global order by aligning spins. Analogously, increasing the bond occupation probability $p$ promotes connectivity and long-range order in percolation. Figs~(\ref{fig:op})  show $P(p,L)$ versus $p$ for different system sizes, demonstrating that $P$ increases monotonically with $p$. This correspondence naturally identifies $p$ as playing a role analogous to an ordering field in percolation.

\subsection{Finite-size scaling and Critical exponents}

Determining the critical exponents of the order parameter, susceptibility, and specific heat—denoted $\beta$, $\gamma$, and $\alpha$, respectively—is a central goal of this work. These exponents characterize the singular behavior of observables near a continuous phase transition and define the universality class. While analytic methods yield their values in the thermodynamic limit, experiments and simulations are limited to finite systems. This limitation is overcome by finite-size scaling (FSS), which relates the behavior of finite systems to asymptotic critical behavior. Near $p_c$, divergences in quantities such as susceptibility and specific heat are truncated by system size, producing rounded and shifted peaks whose scaling with $L$ encodes the underlying singularities. FSS thus provides a systematic route to extract critical exponents from finite-size data.  

The theoretical basis of FSS follows from Buckingham’s $\Pi$-theorem \cite{ref.banerjee2019}. Let $F(p,L)$ be an observable depending on the control parameter $p$ and linear system size $L$. Near the critical point $p_c$, the relevant scaling fields are $(p-p_c)$ and $L^{-1/\nu}$, which share the same scaling dimension. Hence, close to criticality, $(p-p_c)$ can be absorbed into $L$, giving
\begin{equation}
F(p,L) \sim F(L),
\end{equation}
with
\begin{equation}
\label{eq:power_law}
F(L) \sim L^{a/\nu},
\end{equation}
where $a$ is the critical exponent associated with $F$.

This dimensional analysis naturally leads to the dimensionless scaling variable
\begin{equation}
\xi = (p-p_c)L^{1/\nu},
\end{equation}
and the rescaled observable
\begin{equation}
\phi = F L^{-a/\nu}.
\end{equation}
Since $\phi$ is dimensionless, it may depend only on the dimensionless variable $\xi$, yielding the finite-size scaling form
\begin{equation}
\label{eq:fss_c}
F(p,L) \sim L^{a/\nu}\phi_F\left((p-p_c)L^{1/\nu}\right),
\end{equation}
where $\phi_F$ is a universal scaling function characteristic of the observable $F$ \cite{ref.fss_1, Stanley1999, ref.fss_3}.

The order parameter in percolation was also first introduced by Fisher and Essam in 1961 and they defined it as the relative size of largest cluster $P(t,N)=S_{{\rm max}}/N$ which is found to behave like magnetization in magnetic systems \cite{ref.essam1980}. Later, Kasteleyn and Fortuin (1969–1972), through the random cluster model, showed that percolation corresponds to the $q \to 1$ limit of the Potts model \cite{KF1969, FK1972}. These contributions firmly established $P(p,L)$ as the standard order parameter in percolation theory. Order parameter $P$ of percolation takes the typical sigmoidal shape if we plot it
as a function of external field which is in this case is the occupation probability $p$.

The order parameter $P(p,L)$ is known to obey the finite-size scaling form
\begin{equation}
P(p,L) \sim L^{-\beta/\nu}\,\phi_P\!\left ((p-p_c)L^{1/\nu}\right ),
\end{equation}
where $\phi_P(x)$ is a universal scaling function. Since the correlation-length
exponent $\nu$ has already been independently determined from the spanning
probability for each value of the porosity parameter $q$, this scaling relation
can be directly tested numerically. In particular, by plotting
$P(p,L)L^{\beta/\nu}$ as a function of $(p-p_c)L^{1/\nu}$, we obtain the optimal
estimates $\beta/\nu = 0.092$ for $q=0.90$ and $\beta/\nu = 0.075$ for $q=0.85$.
The resulting curves exhibit an excellent collapse onto a single master curve as seen Figs. (\ref{fig:op_1_collapsed}, \ref{fig:op_2_collapsed}),
thereby confirming the validity of the finite-size scaling ansatz.

It is worth emphasizing that the scaling variables
$P(p,L)L^{\beta/\nu}$ and $(p-p_c)L^{1/\nu}$ are both dimensionless. This
immediately implies the asymptotic relations $P \sim L^{-\beta/\nu}$ at
criticality and $(p-p_c) \sim L^{-1/\nu}$ in the vicinity of the transition.
Combining these results leads to the standard power-law behavior
\begin{equation}
P(p,L) \sim (p-p_c)^{\beta},
\end{equation}
from which the critical exponent $\beta$ can be extracted using the
independently determined values of $\beta/\nu$ and $1/\nu$. This behavior is
directly analogous to the scaling of magnetization near the
paramagnetic-ferromagnetic phase transition, reinforcing the interpretation
of $P$ as the appropriate order parameter for the percolation transition. Such high-quality data collapse provides a stringent consistency check and serves as a decisive validation of the extracted critical exponents, confirming that the observed scaling faithfully represents the thermodynamic-limit behavior.


\begin{figure}
\centering
\subfloat[]
{
\includegraphics[height=4.2 cm, width=4.2 cm, clip=true]
{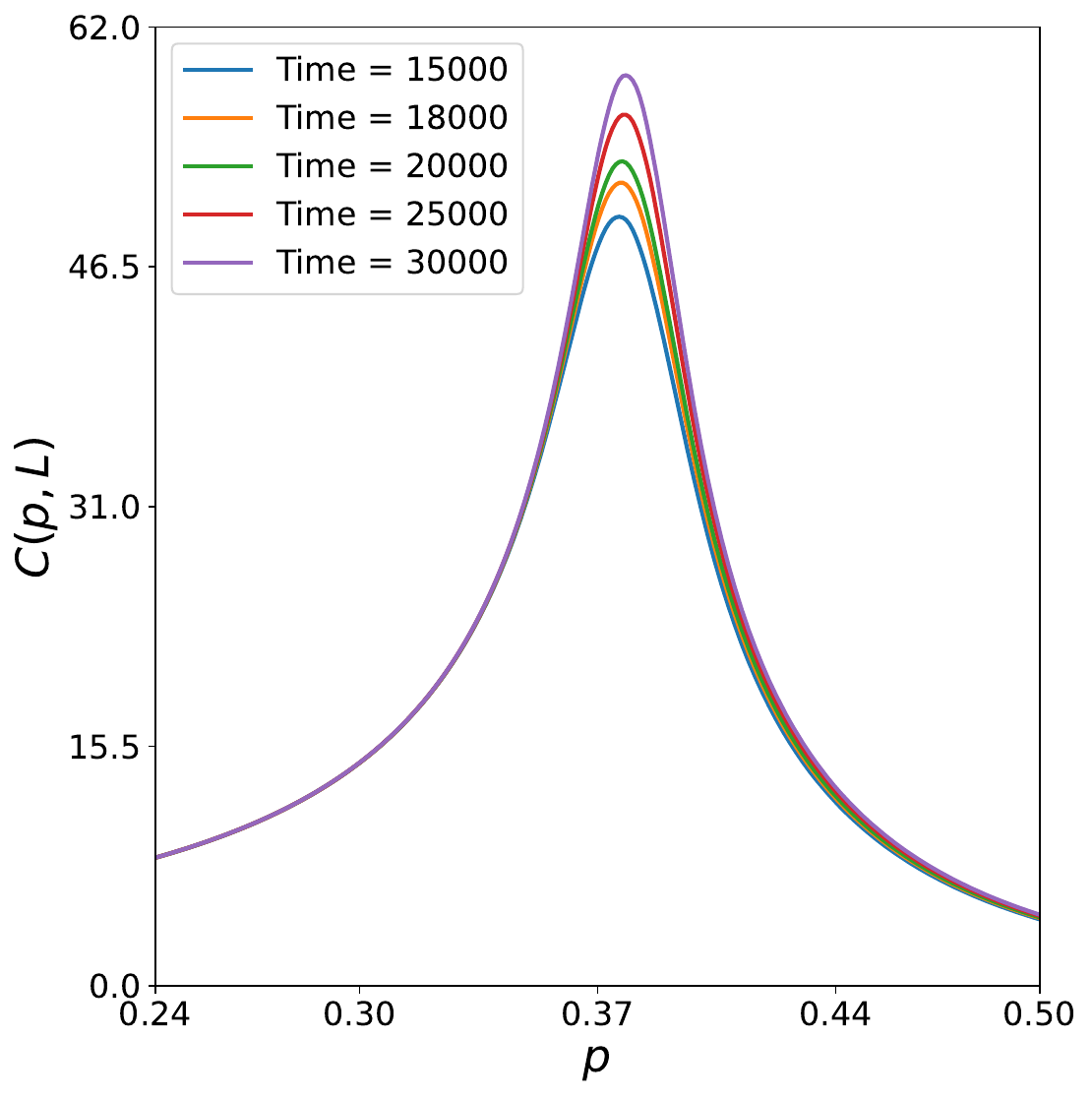}
\label{fig:sp_1}
}
\subfloat[]
{
\includegraphics[height=4.2 cm, width=4.2 cm, clip=true]
{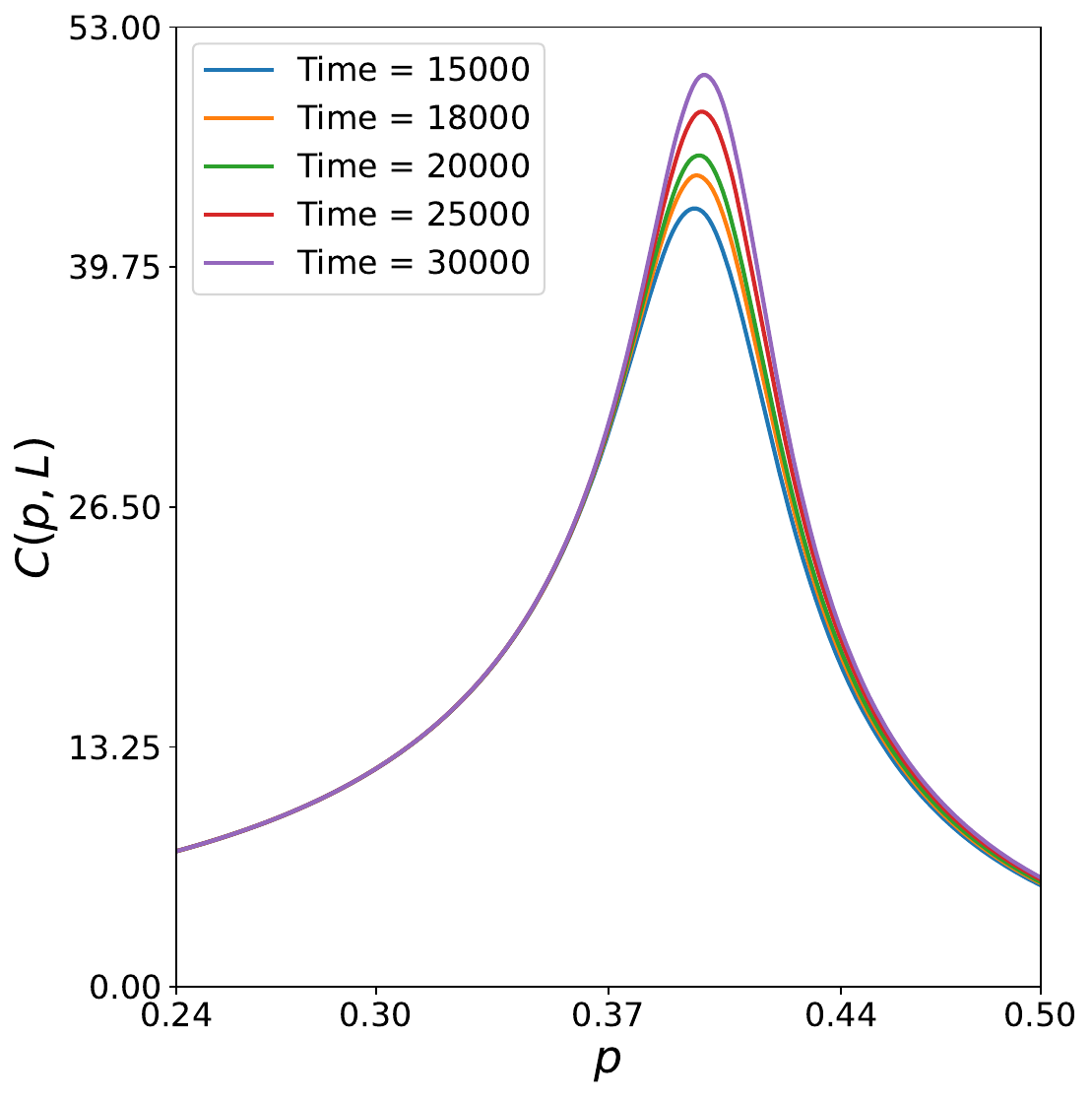}
\label{fig:sp_2}
}

\subfloat[]
{
\includegraphics[height=4.2 cm, width=4.2 cm, clip=true]
{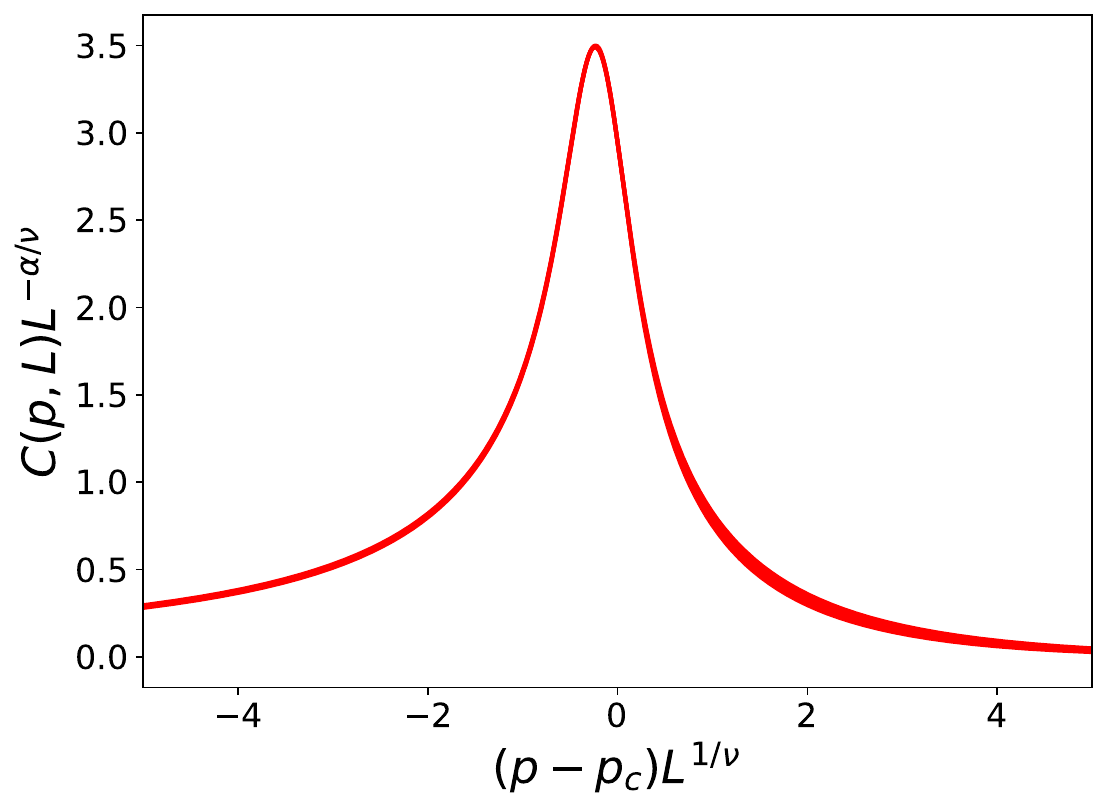}
\label{fig:sp_collapsed_1}
}
\subfloat[]
{
\includegraphics[height=4.2 cm, width=4.2 cm, clip=true]
{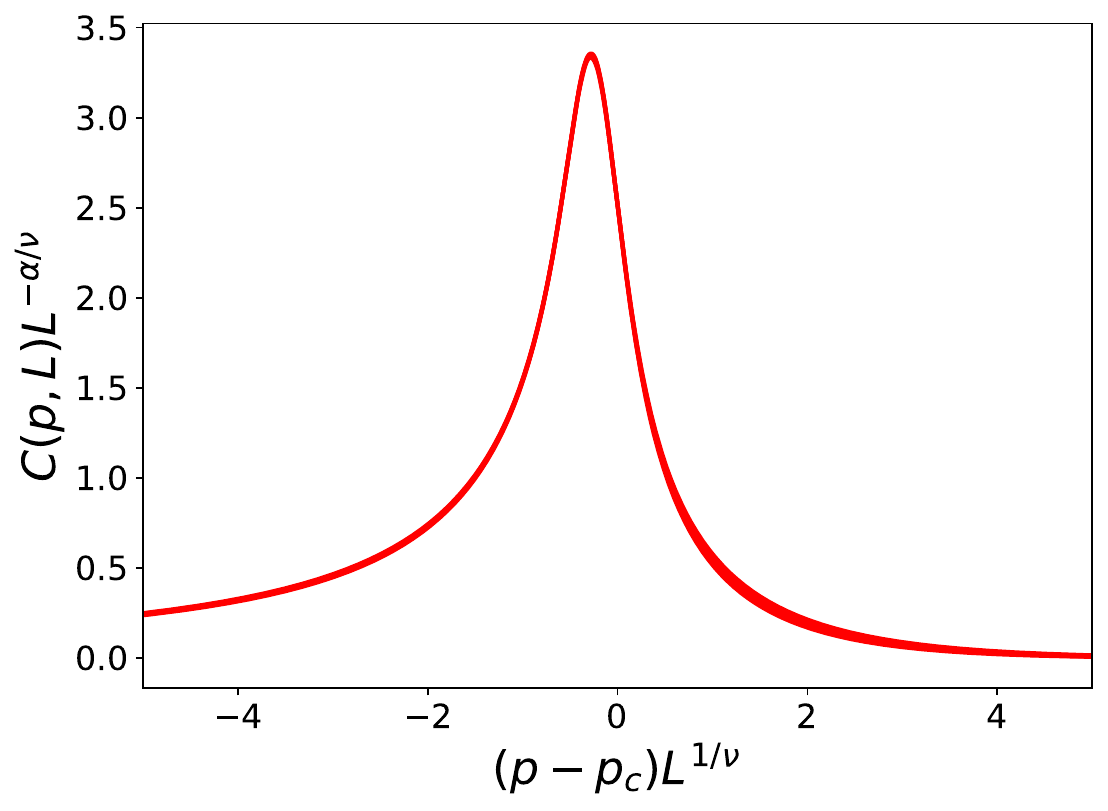}
\label{fig:sp_collapsed_2}
}

\caption{Plots of the specific heat $C(p,L)$ as a function of $p$ for different system sizes are shown in (a) for $q=0.90$ and in (b) for $q=0.85$, illustrating the effect of porosity. Panels (c) and (d) show the corresponding finite-size scaling plots, $C(p,L)L^{-\alpha/\nu}$ versus $(p-p_c)L^{1/\nu}$, which exhibit an excellent data collapse. This confirms that the numerically obtained values of the critical exponents $\alpha/\nu$ and $1/\nu$ are optimal.
} 
\label{fig:sp}
\end{figure}

Using the definition in Eq.~(\ref{eq:sp}), we measured the specific heat by employing the Shannon entropy $H$. Note the analogy with the order parameter, which quantifies the likelihood that a randomly chosen site belongs to the largest cluster. Similarly, Shannon entropy captures the uncertainty or ignorance per attempt. Replacing $S$ with $H$ in Eq.~(\ref{eq:sp}) gives the specific heat for percolation, shown in Figs.~(\ref{fig:sp}) as a function of $p$ for $q=0.90$ and $q=0.85$. To determine the corresponding exponent $\alpha$, we use the finite-size scaling (FSS) hypothesis:
\begin{equation}
\label{eq:fss_specific}
C(p,L)\sim L^{\alpha/\nu}\,\phi_C\big((p-p_c)L^{1/\nu}\big),
\end{equation}
where $\phi_C$ is the universal scaling function for specific heat. Following the same procedure as for the order parameter, we calculate $\alpha/\nu$ for different values of $q$. Since $\nu$ is already known, plotting $C(p,L)L^{-\alpha/\nu}$ versus $(p-p_c)L^{1/\nu}$ results in an excellent data collapse, as shown in Figs.~(\ref{fig:sp_collapsed_1}) and (\ref{fig:sp_collapsed_2}). 
Using the relation $(p-p_c)\sim L^{-1/\nu}$ in $C\sim L^{\alpha/\nu}$, we immediately obtain
\begin{equation}
C(p,L)\sim (p-p_c)^{-\alpha}.
\end{equation}
The quality of the data collapse serves as a consistency check for the accuracy of the estimated exponent $\alpha$. The values of $\alpha$ for different $q$ are listed in the table. This relation demonstrates that, analogous to continuous thermal phase transitions, the specific heat in percolation diverges near the critical point following a power-law.

Following the generating-function formulations of Fisher and Essam~\cite{FE1961} and the considerations of Kasteleyn and Fortuin~\cite{KF1969,FK1972}, one may define a percolation analogue of the free energy as the \emph{number of clusters per lattice site}, corresponding to the zeroth moment of the cluster-size distribution. Explicitly, this quantity is given by
\begin{equation}
n(p) = \sum_{s=1}^{N} n_s(p),
\label{eq:cluster_dens}
\end{equation}
where $n_s(p)$ denotes the number density of finite clusters of size $s$ at occupation probability $p$. The second derivative of $n(p)$ with respect to $p$ is often identified as the analogue of the specific heat. On the square lattice this quantity yields a negative critical exponent $\alpha$, consistent with known results. 

In contrast, the second moment of the cluster-size distribution defines the susceptibility, which measures the expected size of the cluster to which a randomly chosen site belongs. This quantity can be written as
\begin{equation}
S(p) \propto \sum_{s=1}^{N} s^2 n_s(p),
\label{eq:secondmoment}
\end{equation}
and diverges as the percolation threshold is approached from below according to the power law
\begin{equation}
S(p) \sim (p_c - p)^{-\gamma},
\end{equation}
in direct analogy with the divergence of thermodynamic susceptibility near a continuous phase transition. These formulations apply only to finite clusters. Accordingly, in the supercritical regime, the spanning component must be excluded from the definition of $\chi$ in Eq.~\ref{eq:secondmoment}; otherwise, the susceptibility remains trivially divergent. The resulting exponent $\gamma$ is so large that it strongly violates the Rushbrooke scaling relation, which is expected to hold as an equality.

\begin{figure}
\centering
\subfloat[]
{
\includegraphics[height=4.0 cm, width=4.0 cm, clip=true]
{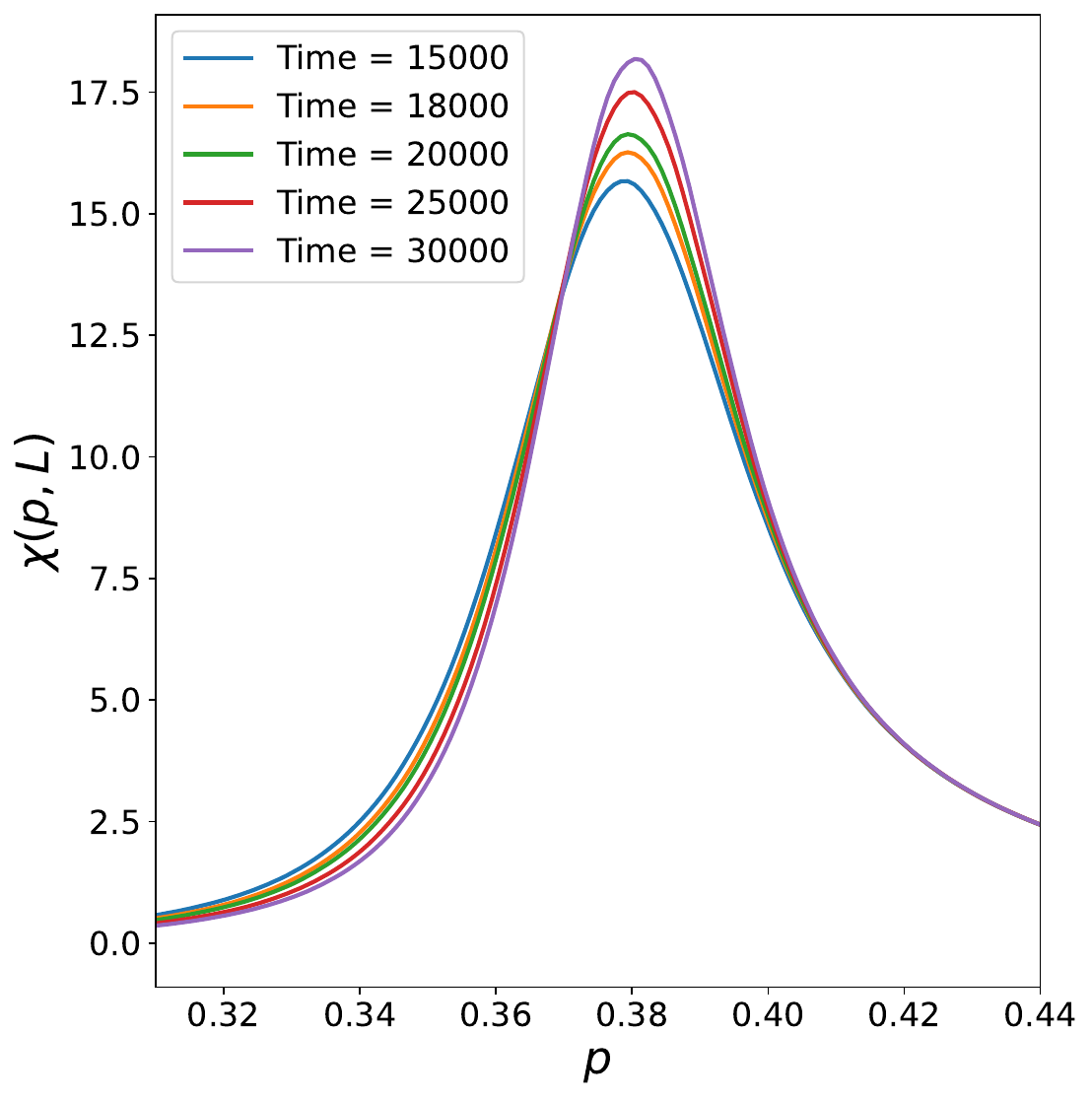}
\label{fig:sus_1}
}
\subfloat[]
{
\includegraphics[height=4.0 cm, width=4.0 cm, clip=true]
{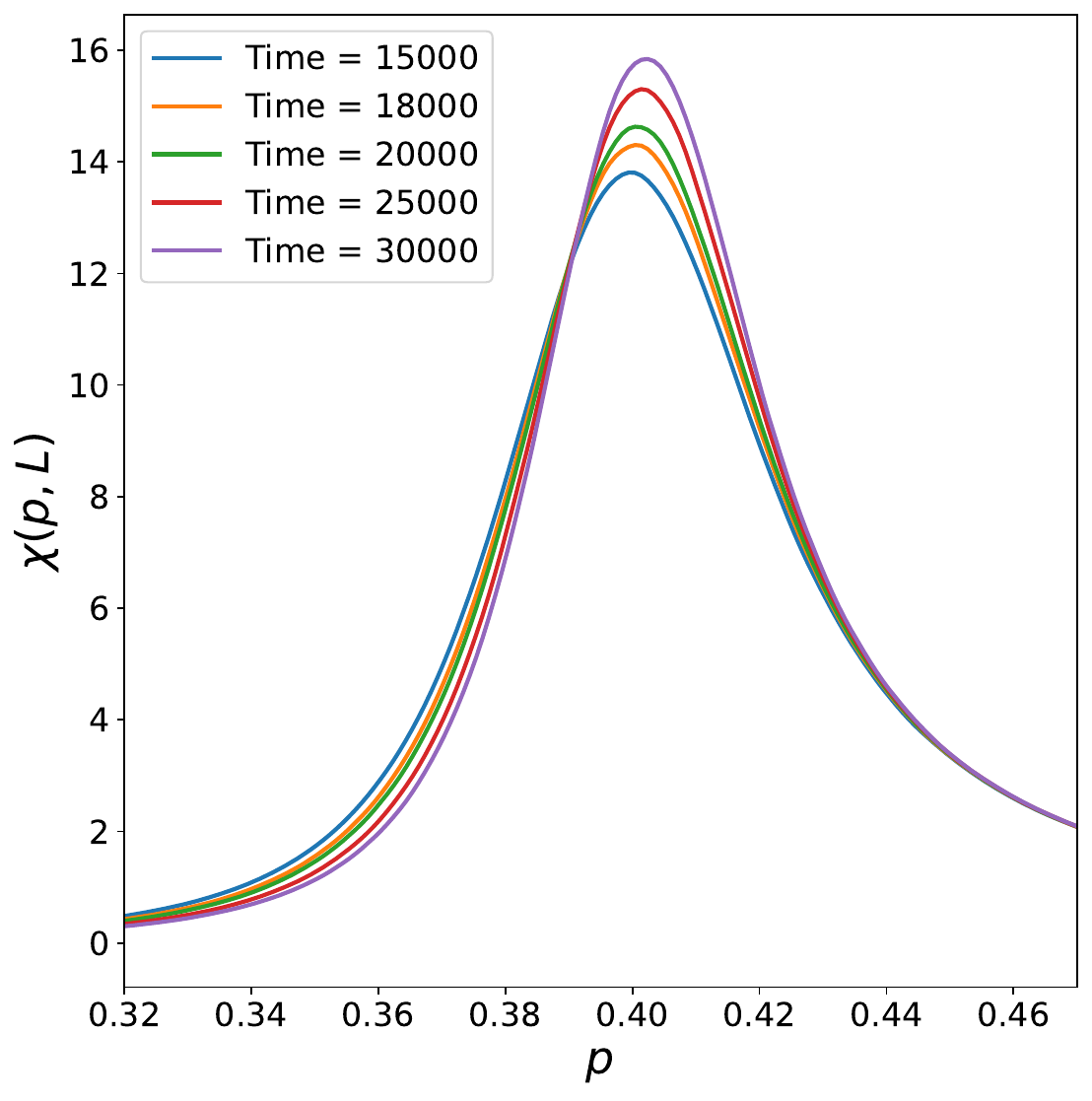}
\label{fig:sus_2}
}

\subfloat[]
{
\includegraphics[height=4.0 cm, width=4.0 cm, clip=true]
{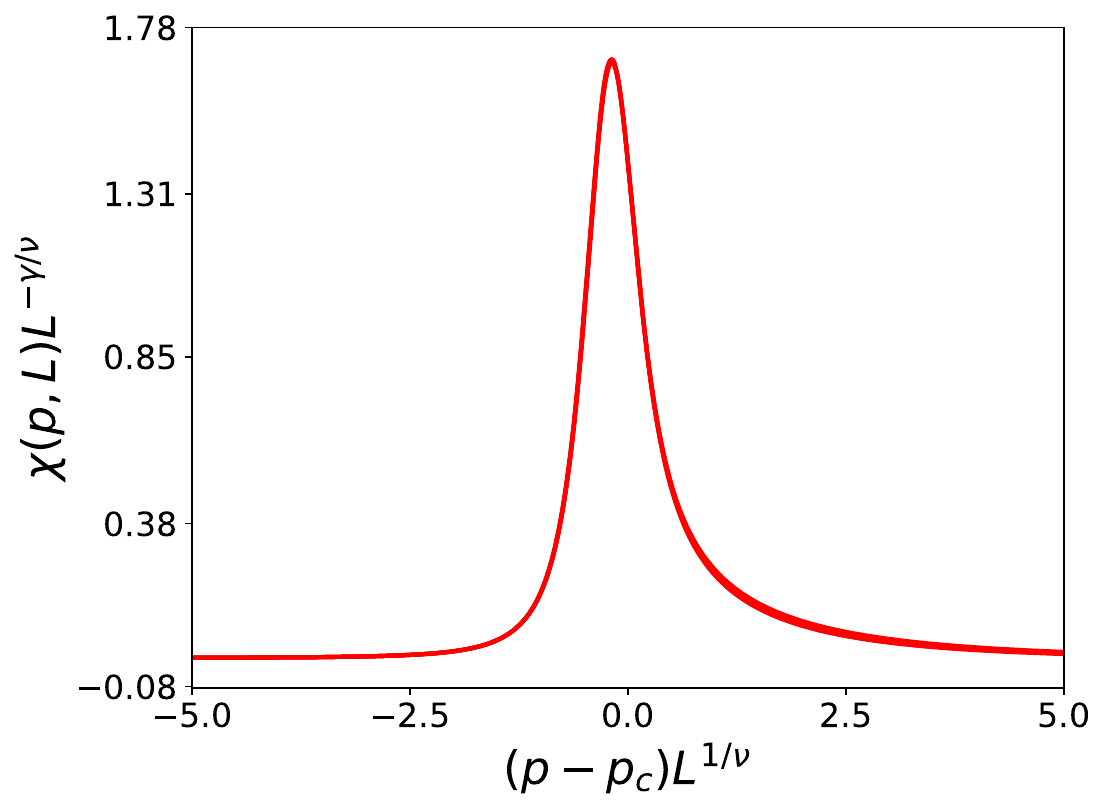}
\label{fig:sus_collapsed_1}
}
\subfloat[]
{
\includegraphics[height=4.0 cm, width=4.0 cm, clip=true]
{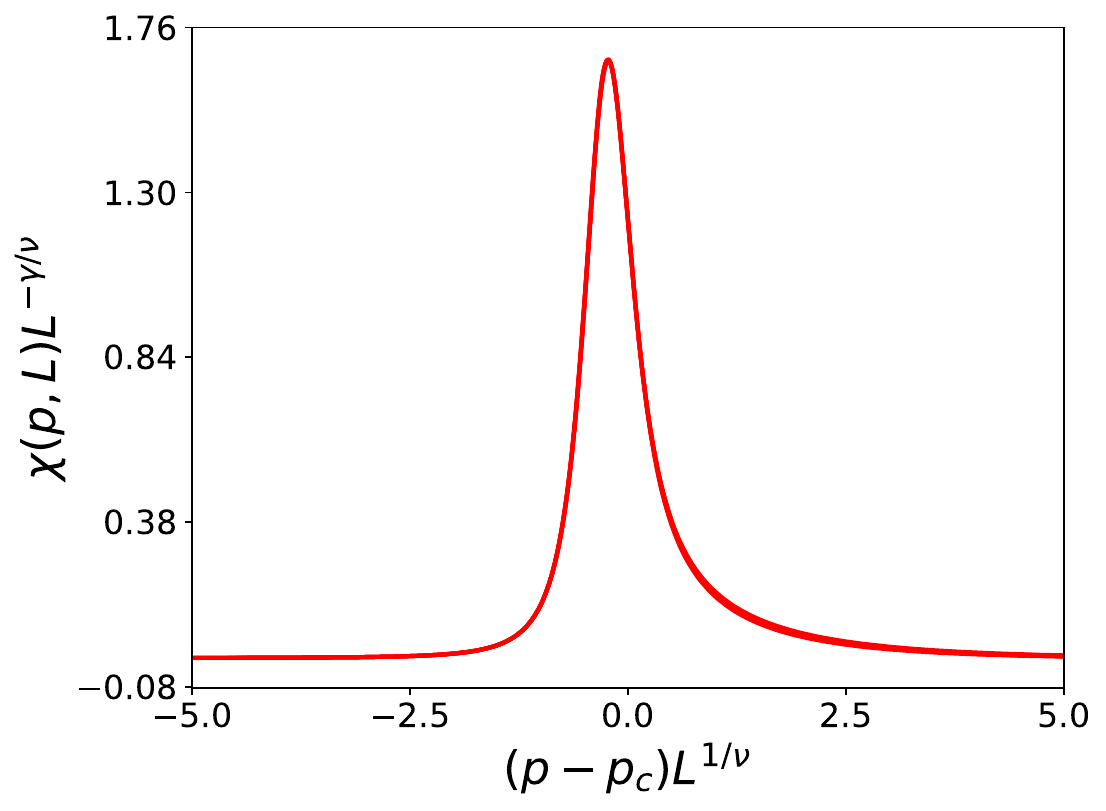}
\label{fig:sus_collapsed_2}
}
\caption{Plots of the susceptibility $\chi$ as a function of the occupation probability $p$ for different system sizes are shown in (a) for $q=0.90$ and in (b) for $q=0.85$. Panels (c) and (d) display the corresponding finite-size scaling plots of $\chi(p,L)L^{-\gamma/\nu}$ versus $(p-p_c(L))L^{1/\nu}$. The excellent collapse of the data indicates that the estimated values of the critical exponent ratios $\gamma/\nu$ and $1/\nu$ are consistent with their theoretical expectations.} 
\label{fig:susceptibility}
\end{figure}

These issues motivated Hassan {\it et al.}~\cite{hassan2017entropy} to redefine susceptibility directly as the derivative of the order parameter with respect to the control parameter $p$, yielding
\begin{equation}
\chi(p)=\frac{dP(p)}{dp}.
\label{eq:susceptibility}
\end{equation}
This definition is thermodynamically consistent and resolves the shortcomings of the conventional second-moment definition: it diverges only at $p_c$ and satisfies the Rushbrooke inequality with near equality for square and weighted planar stochastic lattices~\cite{hassan2017entropy}. The same formulation was subsequently shown to hold for random and competitive percolation models on evolving graphs~\cite{sabbir2018}.  We adopt this definition and compute $\chi(p,L)$ for different $q$, as shown in Figs.~(\ref{fig:sus_1}) and (\ref{fig:sus_2}) for $q=0.9$ and $0.85$, respectively. The behavior closely parallels that of the specific heat. Applying finite-size scaling, we plot $\chi(p,L)L^{-\gamma/\nu}$ versus $(p-p_c)L^{1/\nu}$ and obtain an excellent data collapse (Figs.~(\ref{fig:sus_collapsed_1}) and (\ref{fig:sus_collapsed_2})). This yields the critical divergence
\begin{equation}
\chi(p)\sim |p-p_c|^{-\gamma},
\end{equation}
fully analogous to the divergence of magnetic susceptibility at a continuous phase transition.

\begin{table}
\begin{tabular}{| c | c | c | c | c | c| c } 
\hline  
\ & \ & \ & \ & \ &  \ \ \\[-1mm] 
\ \ WPSPL \ \ & $p_c$ & $\beta$ & $\gamma$ & $\alpha$ & $\alpha+2\beta+\gamma$\\[1mm]  \hline 
\ & \ & \ & \ & \ & \\[-2mm]
q=1.00 & 0.3474 & 0.194 & \ 0.757 \ & \ 0.861  &  2.007 \ \ \\ \hline
\ & \ & \ & \ & \ & \\[-2mm]
q=0.90 & 0.3894 & 0.170 & \ 0.782 \ & \ 0.931  &  2.053	 \ \ \\ \hline
\ & \ & \ & \ & \ & \\[-2mm]
q=0.85 & 0.4165 & 0.155 & 0.809 & \ \ \  0.977 \ & 2.091 \ \ \\
\hline 
\end{tabular}
\caption{Expected values of critical properties when susceptibility is measured by $S(t)$, Eq.~\ref{eq:secondmoment}, with $S(t) \sim (t_c-t)^{-\Gamma}$ and finite size scaling with system size as shown in Eq.~\ref{eq:fss_1}, in addition to the values we determine for IPR via this method. 
}
\label{table:knownvalues}
\end{table}

\section{Summary and Discussion}

In this article, we have investigated bond percolation on the weighted planar stochastic porous lattice (WPSPL), a geometrically disordered substrate whose porosity is controlled by the parameter $q$. Despite its stochastic and porous nature, the WPSPL exhibits robust statistical self-similarity across different growth stages. This property renders the lattice particularly well suited for percolation studies and allows a meaningful comparison with conventional regular lattices. The WPSPL is characterized by several nontrivial structural features: it is locally multifractal, its global (Hausdorff) dimension depends continuously on $q$, and its coordination number distribution is scale free, following a power-law form. These properties place the WPSPL outside the class of standard Euclidean lattices and suggest, a priori, the possibility of unconventional critical behavior.

By determining the percolation threshold and computing the order parameter and entropy, we constructed thermodynamically consistent analogs of temperature and external field for this intrinsically nonthermal system. Substituting these analogs into the definitions of specific heat and susceptibility enabled us to extract the critical exponents $\alpha$, $\beta$, and $\gamma$ using finite-size scaling analysis. Our results demonstrate that the critical behavior of percolation on the WPSPL depends sensitively on the porosity parameter $q$. In particular, we find that decreasing $q$ leads to an increase in the specific-heat exponent $\alpha$ and the susceptibility exponent $\gamma$, while the order-parameter exponent $\beta$ decreases systematically. This continuous variation of critical exponents indicates that, unlike conventional two-dimensional lattices, the WPSPL does not belong to a single universality class.

This finding extends our earlier result that the nonporous weighted planar stochastic lattice ($q=1$) already lies outside the standard universality class of two-dimensional percolation. The present study shows that introducing porosity further enriches the critical behavior by coupling geometric disorder, multifractality, and scale-free connectivity to the percolation transition. Despite the observed nonuniversality, an important consistency check is provided by the Rushbrooke inequality, $\alpha + 2\beta + \gamma \ge 2$, which follows from general scaling arguments. For all values of $q$ studied, this inequality is satisfied, and the sum remains very close to $2$, approaching equality as expected under the static scaling hypothesis \cite{stanleyBook}. This confirms that, although the universality class varies continuously with $q$, the underlying scaling framework remains intact.

Taken together, our results highlight the central role of lattice geometry and global dimension in determining critical behavior. The WPSPL provides a compelling example of a system in which geometric tuning leads to continuously varying critical exponents without violating fundamental scaling relations. This opens new avenues for exploring percolation and critical phenomena on complex substrates that interpolate between disordered, scale-free, self-similar, and multifractal lattices and their porous counterparts.

\bibliographystyle{unsrt}
\bibliography{percolation_porous}

\end{document}